%% file: main.tex
\documentclass[11pt]{article}
\usepackage[a4paper, total={6in, 8in}]{geometry}
\usepackage{graphicx}
\usepackage{natbib}
\usepackage{amsmath}
\usepackage{orcidlink}
\usepackage{amssymb}
\usepackage{amsfonts}
\usepackage{eurosym}
\usepackage{appendix}
\usepackage{authblk}
\usepackage[boxed]{algorithm2e}
\usepackage{booktabs}
\usepackage{enumitem}
\usepackage{tikz}
\usepackage{hyperref}
\usetikzlibrary{shapes.geometric, arrows.meta, positioning, fit}
\usepackage{xcolor}

\definecolor{lightgreenn}{HTML}{D9EAD3}
\definecolor{darkbluee}{HTML}{073763}

\usepackage{caption3} 
\DeclareCaptionOption{parskip}[]{} 
\usepackage[footnotesize]{caption}

\usepackage{doi}

\providecommand{\keywords}[1]
{
  \small	
  \textit{{Keywords --}} #1
}
\providecommand{\highlight}[2]{%
  \small
  \textit{{Highlights }   } { #1}
  \begin{itemize}[leftmargin=1.4cm, nosep]
    #2
  \end{itemize}
}

\title{
    Evaluating Impacts of Traffic Regulations in Complex Mobility Systems Using Scenario-Based Simulations
}
\author{Arianna Burzacchi$^{1*}$\orcidlink{0000-0001-8284-4909} and Marco Pistore$^{1}$\orcidlink{0000-0003-1425-942X}  \\
        \small $^{1}$ Modeling and Simulation of Socio-Technical Systems, \\
        Fondazione Bruno Kessler, Via Sommarive 18, 38123, Trento, Italy \\
        \small $^{*}$ Corresponding author: aburzacchi@fbk.eu \\
}

\date{}

\begin{document}

\maketitle

\begin{abstract}
Urban traffic regulation policies are increasingly used to address congestion, emissions, and accessibility in cities, yet their impacts are difficult to assess due to the socio-technical complexity of urban mobility systems. 
Recent advances in data availability and computational power enable new forms of model-driven, simulation-based decision support for transportation policy design. 
This paper proposes a novel simulation paradigm for the ex-ante evaluation of direct and indirect impacts, spanning traffic conditions, transportation-related effects and economic accessibility. 
The approach integrates a multi-layer urban mobility model combining a physical layer of mobility flows and emissions with a social layer capturing behavioral responses and adaptation to policy changes. Real-world data are used to instantiate the current as-is scenario, while policy alternatives and behavioral assumptions are encoded as model parameters to generate multiple what-if scenarios. The framework supports systematic comparison across scenarios by analyzing variations in simulated outcomes induced by policy interventions. 
The proposed approach is illustrated through a case study that aims to assess the impacts of the introduction of broad urban traffic restriction schemes. Results demonstrate the framework’s ability to explore alternative regulatory designs and user responses, supporting informed and anticipatory evaluation of urban traffic policies.
\end{abstract}

\keywords{Urban mobility; traffic regulatios; what-if scenario; modeling complex systems}

\section{Introduction} \label{sec:intro}

The availability of data for sensing and observing urban systems has grown at an unprecedented pace. Advances in pervasive sensing, mobile devices, connected vehicles, and digital platforms now provide access to urban data with increasingly fine-grained spatial, temporal, and social resolution \citep{zheng2019urban}. In parallel, rapid progress in computational capabilities has substantially expanded. Together, these developments are enabling public authorities and urban decision makers to adopt novel, powerful data-driven paradigms, allowing the systematic evaluation of policy interventions beyond what was previously feasible.

These technological advances have brought renewed attention to one of the central research frontiers in urban studies and transportation science: complexity. Cities cannot be adequately understood as collections of independent subsystems or through purely reductionist models. Rather, they exhibit the defining characteristics of \textit{complex systems-of-systems}, in which multiple interacting subsystems, such as transportation, land use, energy, governance, and social behavior, co-evolve over time. Moreover, urban systems are inherently \textit{socio-technical}, as technical infrastructures and control mechanisms are tightly coupled with human decision-making, social norms, institutional constraints, and adaptive behaviors. These interactions give rise to non-linear dynamics, feedback loops, path dependencies, and emergent phenomena that challenge traditional modeling and evaluation approaches.

\textit{Urban mobility systems} constitute a paradigmatic example of such socio-technical complex systems \citep{Batty2008The, McPhearson2016Advancing, Luca2024Civic}. Transportation networks are not merely physical infrastructures supporting vehicle flows; they are shaped by---and in turn shape---individual and collective behavior, economic incentives, regulatory frameworks, and social practices. Interventions such as traffic regulations, access restrictions, pricing schemes, or infrastructural changes can trigger individual and social responses and behavioral adaptations that are difficult to anticipate using aggregate or static models alone.

This paper contributes to this line of research by proposing a \textit{model-driven, simulation-based paradigm for decision support in urban mobility policy design}. The objective is to enable the ex-ante evaluation of direct impacts, such as traffic conditions, congestion, modal split, and emissions, and indirect impacts arising in connected domains. By supporting systematic exploration of policy alternatives before implementation, the proposed approach aims to reduce uncertainty and unintended consequences in urban traffic regulation. 

The approach is inherently \textit{simulation-based}, allowing the representation of the current \textit{as-is} state of the mobility system alongside multiple \textit{what-if scenarios} corresponding to alternative traffic policies and structural changes in urban mobility. This enables controlled experimentation on policy options that would be impractical, costly, or ethically problematic to test in real-world settings.

At its core, the framework relies on an \textit{urban mobility model} that explicitly integrates a physical layer, including the mobility traffic flows and pollutant emissions, and a social layer capturing individual and collective behavior, adaptation mechanisms, and varying degrees of behavioral rigidity. This coupling is essential to represent the socio-technical nature of urban mobility systems and to account for behavioral responses to policy interventions.

The as-is scenario is instantiated using directly measured and estimated quantities derived from available data sources. These include mobility-related indicators such as vehicle volumes, Origin–Destination matrices, and emissions, represented at both macro and micro levels. Policy alternatives are then defined through model parameters that encode endogenous events (e.g., public authority regulations), exogenous events (e.g., infrastructure disruptions, changes in vehicle fleets), and assumptions on user responses and behavioral adaptation.
Simulation outputs characterize the system behavior under each of these what-if scenarios, with particular emphasis on variations in outcomes such as traffic patterns and emissions, induced by changes in model parameters. The framework also supports systematic comparison across scenarios and against the baseline as-is condition. 
We remark that the scope of this work is not causal inference or factor discovery: we assume that relevant factors and their relationships to the modeled phenomena are given and encoded in the model structure, and policy analysis is conducted by manipulating input parameters, generating alternative system states, and comparing their simulated outcomes.

To illustrate the proposed approach, we apply it to a \textit{case study in the city of Bologna}, Italy, a major urban center characterized by a large resident population and a complex, multimodal mobility system. The case study focuses on assessing the direct and indirect impacts of a hypothetical newly introduced traffic policy aimed at regulating vehicular circulation within a designated urban area. Building on the city’s existing limited traffic zone in the historical center, we simulate an additional circulation restriction applied to a broader area. This setting enables us to demonstrate the ability of the proposed framework to construct, analyze, and compare multiple what-if scenarios, reflecting alternative regulatory designs and heterogeneous user responses, and to support informed evaluation of traffic regulation policies before their implementation.

One key aspect of the proposed simulation approach is the attention given to uncertainty. This includes recognizing that variability in input data, modeling assumptions, and behavioral responses can influence both as-is and what-if scenario outcomes, and, where possible, quantifying its impact. While this aspect will not be the primary focus of the paper, the fact that the framework intrinsically supports it paves the way for future extensions toward fully uncertainty-aware evaluations.

This paper is structured as follows. Section \ref{sec:conceptual} presents the conceptual framework of the study, highlighting the transportation policies and the standard evaluation methods, as well as explaining the challenges in modeling mobility systems. Section \ref{sec:meth} outlines the general architecture of the proposed methodological pipeline and its application to the specific case study of the city of Bologna. Section \ref{sec:math} formalizes the mathematical model constructed for the case study. Section \ref{sec:application} details the implementation and application of the model, reporting the assessed results. Finally, Section \ref{sec:conclusions} summarizes the main findings and discusses the limitations and challenges for future research.

\section{Conceptual framework}\label{sec:conceptual}

\subsection{Policy application and evaluation}

In recent decades, transportation policies have assumed a central role in the international political agenda, driven by the need to balance the traffic externalities with economic growth, social equity, accessibility, and environmental objectives \citep{Lindsey2020Addressing}. The literature has identified various types of traffic policy interventions, reflecting different strategic approaches to address transport challenges \citep{Santos2010Pt1, Santos2010Pt2}. Operationally, these policies can be categorized according to whether they target the supply side or the demand side of the transport system.

Supply strategies are designed to increase the quality of the supply and, hence, its traffic capacity and efficiency. Common examples of such strategies include upgrading infrastructure, implementing smart technology for real-time traffic updates, and refining public transport schedules to better match commuter needs. As demonstrated by \citet{Rhoads2006Why}, supply-side policies tend to have a higher social discount rate and are often preferred to obtain immediate and tangible changes in the transport system. However, they may induce additional demand and generate the so-called induced traffic, sometimes offsetting long-term benefits \citep{Goodwin1996Empirical, Downs2004Still, Hymel2019If}.
Alongside, demand-side traffic policies address the mitigation of congestion by altering the demand for transportation \citep{Lindsey2023Distributional, Small2024Economics}. These policies take many forms, for instance: restricting vehicle access in certain zones, using taxes to discourage car ownership, adjusting public transport fares, and implementing pricing schemes to reduce traffic congestion. 

As widely described by \citet{Depalma2011Traffic}, the enactment of pricing policies serves to internalize the marginal social cost imposed by travelers: who contributes to collective traffic congestion and its negative externalities is required to bear their share of the collective cost. Beyond the original theoretical framework for computing optimal traffic fees \citep{Pigou1920Economics}, practical implementations of pricing policies exhibit considerable variation. Common approaches include area-based charging (i.e., tolls for entering or passing through designated zones) and facility-based charging (i.e., tolls for using specific roads or lanes), with fees that vary from flat daily rates to advanced real-time pricing \citep{Bonsall2007Responses}.
Pricing schemes commonly differ by vehicle characteristics (e.g., fuel type, registration year, weight) and may include discounts or exemptions for specific vehicle categories and driver groups. 

The implementation of these traffic policies raises the critical question of how to evaluate their effectiveness and measure their impact on congestion mitigation. Many studies have focused on assessing the effects of traffic policies on urban factors of interest, from traffic itself to economic growth and social equity, utilizing several analytical tools such as quantitative descriptions, causal inference, and econometrics to evaluate ex-post effects \citep{Foreman2016Crossing, Zhang2025Ex, Gibson2015Effects, Borjesson2018Swedish}. 
Ex-post evaluations as such, however, require the policy to be enacted before its assessment is possible, and hence fail in providing policy makers with a formal and quantitative analysis of the short- and long-term effects of the policy. Instead, ex-ante evaluations address this task and provide evidence on the effectiveness, or not, of a policy proposed. 

Cost-Benefit Analysis (CBA) represents the standard tool for ex-ante policy evaluation, being among the most widely discussed, tested, and used techniques \citep{Koopmans2020Chapter1}. In CBA, the societal value of a new policy is evaluated by comparing in monetary terms the total cost related to its implementation (e.g., planning and construction costs) and the benefits that it causes (e.g., savings in fuel consumption, savings in maintenance).
While it is effective at assessing direct and tangible effects, CBA is often criticized for its difficulty in fully capturing externalities, which are not easily linked to a clearly defined economic metric and can result in distorted or biased estimates \citep{Naes2020Chapter9}.
Other comprehensive methodologies have emerged in response to the persistent gap between standard policy evaluation practices and the theoretically solutions \citep{Oliveira2010Evaluation, Koopmans2020Chapter1, Naes2020Chapter9}. However, a major challenge remains: once the quantities of interest have been identified, one must propose models that can capture the system's response mechanisms to policy implementation and adequately adjust these quantities accordingly. The adoption of such modeling approaches is indeeed necessary to estimate outcome differences rigorously and robustly, thereby allowing for their appropriate evaluation.

\subsection{Socio-technical mobility systems: models and challenges} 

Since their initial development in the 1950s, the goal of mobility models has been to provide a systematic methodology for quantifying the effects of changes in the underlying transportation systems \citep{Vannes2020Chapter4}. For instance, the four-step model is frequently used to forecast the impact of major infrastructure projects \citep{McNally2007Four}, while System Dynamics serves as a sample approach for exploring how complex long-term policies might evolve \citep{Shepherd2014Review}. Recent advances and new applications now enable researchers to address more specific questions and study diverse transportation contexts.
Macroscopic models capture aggregate system behavior \citep{ Wilson1970Entropy, McNally2007Four, Shepherd2014Review}, while microscopic models represent individual agents and their interactions \citep{Transims2001, Horni2016Matsim, Sumo2018, Vissim2025}, and both are widely used to explore what-if scenarios, such as assessing how mobility patterns may change under alternative policy interventions \citep{Rieser2008Modeling, Brand2019Lifestyle, Melkonyan2020Integrated, Daniel2021City, Asgarpour2023Infrastructure}.

While models differ in focus and detail, they address and integrate fundamental components to capture the essential elements for describing, understanding, and analyzing the complex phenomena under study. 
First, mobility systems are intrinsically socio-technical complex systems \citep{Batty2008The, McPhearson2016Advancing, Luca2024Civic}. Social components (e.g., people, organizations, cultures, processes) and techniques (e.g., technologies, infrastructures, tools) are deeply interconnected and influence each other. They integrate physical infrastructures, technologies, and digital systems with individual behaviors and social practices components, and the interactions of these closely connected components give rise to complex, emergent behaviors and trade-offs inherent to large-scale socio-technical systems \citep{Sussman2000Introduction}. Capturing this complexity is a key challenge and requires comprehensive, fair, and efficient models capable of supporting and improving informed decision-making processes on policy development and scenario analysis. 

Uncertainty is another inherent feature of mobility systems, closely related to their complexity, as it measures ``any deviation from the unachievable ideal of completely deterministic knowledge of the relevant system" \citep{Walker2003Defining}. Uncertainty emerges across multiple dimensions: data limitations may introduce bias and incompleteness; simplifications and context-dependent assumptions at the modeling stage can obscure complex causal relationships; small inaccuracies in forecasting accumulate over time, potentially leading to large deviations between projected and observed outcomes \citep{Mannucci2023Exploring}. Cities face a wide range of unpredictable events, from natural disasters and health emergencies to sudden economic shifts or policy changes. Even in stable conditions, human behavior adds another layer of unpredictability: people adjust their habits, react differently to incentives, and sometimes resist new regulations in ways that defy precise modeling \citep{Avineri2004Violations}.

Another key challenge is the growing demand for transparency, interpretability, and accountability. Stakeholders require clear explanations of model assumptions, mechanisms, and limitations to trust the results and use them effectively for decision-making. Similarly, citizens need understandable information about how models influence policies and daily life, so that they can engage with decisions that affect their communities \citep{Luca2024Civic}.

The domains of modeling, analysis, and civic engagement are all shaped by these challenges. Rather than treating them as isolated problems to be eliminated, researchers should acknowledge their interdependence and adopt an integrated, holistic approach. By encompassing this interconnected perspective, modelers can create tools that are both analytically rigorous and socially meaningful, supporting trust, participation, and more informed policy choices.

\section{Methodological framework} \label{sec:meth}

\subsection{Motivating case: traffic regulation pricing policy} \label{subsec:moti}

The illustrative case driving this study considers an urban area characterized by recurring congestion during peak hours and significant levels of atmospheric emissions. In such contexts, implementing traffic regulation policies is crucial, yet their evaluation proves challenging due to the complex interactions between user behavior, network characteristics, and regulatory parameters.

In the scope of this paper, the policy under study focuses on a specific regulatory action, namely, the \textit{introduction of a traffic pricing system} where tolls vary according to temporal and categorical criteria, with exemptions for specific user groups and a defined application schedule.
Even with this limited focus, the complexity of the case becomes evident when observing how demand responds to the implemented measures. Users' choices are heterogeneous and influenced by factors such as travel time, availability of alternative transport modes, and individual habits. The overall effect of the policy on traffic and emissions cannot be determined in a purely intuitive manner, as interactions between individual behaviors and regulatory measures generate nonlinear dynamic phenomena. In addition, uncertainty may be present in the unpredictable strategic response of users to the applied regulation. This clearly demonstrates the need for a systemic approach capable of simultaneously modeling physical urban layers, user strategies, and policy parameters. Evaluating the effectiveness of interventions requires a framework that can represent both the policy and the behavioral diversity of users, while preserving the dynamic interdependencies within the system.

The main contribution of this framework is to generate comparable as-is and what-if scenarios, and to provide the analytical baseline for policy appraisal. By producing consistent as-is and what-if scenarios, it ensures that ex-post evaluations are grounded with explicit assumptions and modeling choices, and reproducible methods and procedures. The general framework is introduced in Section~\ref{subsec:gene}, while Section~\ref{subsec:appli} introduces a motivating case: it provides a concrete illustration of the challenges which the proposed framework aims to address, and clarifies why a systematic approach is necessary to link regulatory interventions, behavioral responses, and resulting system states in a coherent and reproducible manner. 

\subsection{General framework architecture} \label{subsec:gene}

\input{schemas/schema1}


Within a spatio-temporal domain, a \textit{scenario} is described by its core quantities of interest, which are the measurable variables that the model aims to consider, describe, and evaluate. They refer to the relevant aspects of reality, including physical and environmental features (such as the road network and the land use), social characteristics (such as the number of people with specific attributes), and the dynamics of the complex civic system. 
In the as-is scenario, the quantities of interest are inferred directly from the real-world data within the spatio-temporal domain of the analysis. Indeed, the as-is scenario represents the current conditions of the analysed framework and serves as a benchmark for measuring the impact of any intervention introduced. In contrast, the what-if scenario represents the situation after the implementation of changes in the current system, and the values of the quantities of interest are estimated throughout the model. 

The changes described above can result from either endogenous or exogenous events. Endogenous events emerge from within the system itself, arising from internal interactions or natural changes that develop over time within the system. Exogenous events originate from outside the system, either being unforeseen (e.g., unexpected changes in the environment or external shocks) or being imposed by external agents (e.g., new regulations). In both cases, the goal is to model these events using a series of \textit{control parameters}. Their definition is problem-dependent and has the objective of identifying the key characteristics of the event and schematizing them with a certain number of parameters. 

When an event alters the current state, changes occur as direct and indirect \textit{consequences of, and reactions to, the event}. Direct effects refer to interventions that impose structural modifications on the system (e.g., closing a road that alters the accessibility of the road network, or constructing new buildings with public services that change land use patterns). Indirect effects include aspects within the transport domain (e.g., the modal shift towards public transport) or strongly correlated (e.g., impacts on logistics), but they also include broader secondary effects, such as impacts on the city economy (e.g., impact on merchants and shopkeepers, impact on city tourism, etc.) and social aspects (e.g., limiting the right to mobility, or imposing disproportionate disadvantages on specific vulnerable groups). The effects are complex to model, as they require identifying the potential behavioral response of people and describing it through explicit formulations.

The model accounts for user responses by identifying and representing plausible \textit{behavioral strategies}. For instance, when new traffic regulations are introduced, users may either maintain their current habits or adjust them, depending on policy relevance, adaptation convenience, or inherent user characteristics. In defining these strategies, it is important to capture the interdependencies between possible behaviors and the features of both the scenario and the policy. Accordingly, strategies are specified conditionally, based on control parameters and additional behavioral factors.

The development of the what-if scenario begins from the original system conditions, as illustrated in Figure \ref{fig:intro:scheme}. These conditions are altered through the occurrence of an event, defined by control parameters, which directly modifies the system and indirectly the users' habits in the system. The combined effects extend through the system, affecting the key variables under study. The \textit{impact evaluation} is hence assessed by comparing the as-is initial measures with the new what-if ones. The extent of the change is measured as the variation in the quantities of interest.

In this framework, variables can be defined either deterministically, with fixed values, or stochastically, as random variables following probability distributions. The probabilistic formulation allows us to explicitly represent input \textit{uncertainty} and trace how it propagates through the model to the final estimates of the quantities of interest. By treating parameters probabilistically rather than as fixed values, the model can be evaluated repeatedly, with each simulation producing potentially different outputs that reflect the underlying uncertainty. An ensemble approach enables the quantification of output variability, transforming single-point forecasts into a range of probable outcomes. This allows for the derivation of confidence intervals that provide a more statistically-rigorous foundation for comparing the as-is and what-if scenarios. 

\subsection{Application to the case study} \label{subsec:appli}

In the motivating case, the primary goal is to assess the effectiveness of a newly introduced regulations concerning vehicular traffic in a specific urban area. Even if the ultimate goal is to build a general framework for traffic policy appraisal, in order to keep the discussion focues, the case is limited a regulation concerning the introduction of a traffic fee meant to limit vehicle circulation. Given this scope, key variables for analysis include vehicular traffic flow measurements alongside additional metrics on vehicle emissions and air quality, which will act as performance metrics for the restrictions on circulation and on pollution. The policy is defined by control parameters that quantify the regulatory restriction across temporal boundaries, affected user bases, and economic burdens imposed by the measure. Furthermore, the model captures user reactions as a function of both behavioral attributes and the defining properties of the policy. This Section presents these variables and the interconnected relationships that allow moving from one to the other, as summarized in Figure \ref{fig:scenario}.

Three core variables are observed and monitored in the as-is and what-if scenarios. Two quantities control the number of vehicles and thus measure the direct effects of policy limitations: the \textit{inflow} represents the number of vehicles entering the area affected by the policy regulation, and \textit{traffic} defines the number of vehicles circulating in the area. Then, the \textit{emissions} measure the overall pollutants generated by the vehicles circulating in the area, sheding lights on the policy impact on health and environment. 

The policy enacts an area-based traffic regulation for which the vehicular circulation in a designated area is limited. In this example, the regulation imposes a fee on all vehicles circulating into the area during the daily hours. The cost depends on the type of vehicle in circulation: to achieve emission reduction goals alongside traffic reduction, different penalty levels can be applied to vehicles based on their emission standards, for instance, with higher costs assigned to more polluting classes. Moreover, the policy provides payment exemptions for vulnerable social groups. 

In response to the implementation of the regulatory policy, affected travelers reassess their transportation habits and may modify them accordingly. The model introduces behavioral parameters, designed to define users’ attitudes and preferences and account for the most likely behavior strategies. Among them, the most plausible, included in the model, are: \textit{rigidity} (i.e., continuing with the usual mobility habits), \textit{time-shifting} (i.e., shifting the trip to an earlier time before the policy becomes active or to a later time after the policy ends), and \textit{mode-shifting} (i.e., changing the mode of transport from private vehicles towards public transport means). If none of the alternatives are viable, the model incorporates the \textit{lost} strategy; this represents  
trip cancellations, for instance, but also changes to other modes of transport not monitored in this model (e.g., foot, bicycle).

The application of the policy and its effects on the user travel behaviors initiate a sequential process that progressively alters inflow, traffic, and emission levels in the \textit{what-if scenario}.
The effects on the inflow follow directly from the policy implementation. In the new scenario, the inflow varies from the base one with time: it may be reduced when the policy is applied and there remain only exempted and rigid travelers; on the other hand, it may be increased in the hours of policy inactivation, when anticipating and postponing travelers are to be added to the base inflow. 
When aggregating traffic from all vehicles, both those starting their journey inside the area and those arriving from outside, the total traffic volume changes in accordance with the variations in the number of vehicles. The new scenario is hence associated with a modified value of total traffic in the area per time instant. This causes changes in the vehicular emissions from the area, which vary according to the traffic.
The policy and the subsequent scenario modification affect several key metrics necessary for a full policy evaluation. Illustrative examples include direct transport metrics, such as the extent of modal shift, and various indirect impacts, such as social mobility constraints and broader economic consequences for tourism and commerce. While not the primary focus of this study, these secondary effects are an important dimension of the phenomenon and are currently being examined in ongoing work, with further analyses planned in future research. 

\input{schemas/schema2}

\section{Mathematical formulation} \label{sec:math}

\subsection{Scenario definition} \label{subsec:scen}

The analysis focuses on an urban area, of which a portion of interest is designated for the policy implementation, and other external areas surround it. The analysis considers one reference working day, with measures taken during the 24 hours of the day with a sampling frequency $\nu$ (e.g., every 5 minutes). 
The policy is assumed to be applied on the reference day during designated time slots, with a starting time $t_s$ and an ending time $t_e$. A fraction of vehicles is exempted from its action, $F_e$, while it targets non-exempt vehicles with the fee amount varying according to the vehicle emission standard: assuming there are $n_l$ standards, the costs are $\{C_{0l}\}_{l=0,\dots,n_l}$.

Among the quantities of interest, $I(t)$ represents the vector of inflows, that is, the number of vehicles entering the area in each time interval $t$. In the as-is base scenario, these inflow values are treated as known parameters, obtained through estimation procedures applied to real traffic data, while in the what-if scenario, their values are estimated with the application of the behavioral models explained in Section \ref{subsec:inflow}. The other two core variables are $T(t)$, representing the number of circulating vehicles in the area during time interval $t$, and $E(t)$, which denotes the emissions produced by all vehicles circulating within the area during time interval $t$. The estimation of these quantities is presented in Section \ref{subsec:traffic} and Section \ref{subsec:emissions}, respectively, and follows from the known inflow patterns combined with the emission level distribution of the active vehicle fleet. 
We remark that the explicit models introduced below provide one possible implementation of the framework. Other specifications can be used within the same framework, which accomodates a wide range of alternative modeling assumptions.

\subsection{Inflow modifications} \label{subsec:inflow}

The implemented policy, adding a fee for vehicle access to the designated area, is expected to have significant effects on the vehicles circulating in the area.
Indeed, the group of travelers directly subject to the fee is very likely to reconsider and adjust their transportation choices. 
The model simulates different behavioral strategies applied by directly affected travelers in response to the policy application: the maintenance of previous habits (i.e., rigidity), the variation of the travel starting time to avoid the time slot of the policy (i.e., time-shifting), the variation of the mode of transport from private vehicles to another transport means (i.e., mode-shifting), and the cancellation of the travel.
These options represent only a subset of the possible responses, chosen as the most relevant for this case study, while acknowledging that other behavioral adaptations may also occur. Moreover, since individuals indirectly affected by the policy, such as those exempt from it or entering the area outside its activation hours, are expected to maintain largely unchanged behavior, the model formulation concentrates exclusively on the vehicle segment directly influenced by the policy.

The evaluation of these strategies reflects the trade-offs between convenience, cost, and compliance with the policy. In general, the likelihood of each behavioral option is evaluated by looking at the difference between the effort required by the policy and the maximum acceptable effort. ``Effort" here refers to the general cost or burden imposed by the policy rather than solely monetary expense. For instance, advancing the departure time will weigh the burden of schedule changes against the willingness to travel earlier; evaluating modal shift will assess the effort to favour alternative transport over private vehicles; while contemplating unchanged behavior will compare the monetary cost imposed by the policy against the willingness to pay. 

The probability of each behavioral response is modeled using distinct formulations, reflecting the aggregate behavioral tendencies observed in similar contexts. Rather than representing individual-level decision processes, these formulations capture the emergent collective behavior resulting from heterogeneous preferences and constraints. Then, the optimal behavior is identified based on the acceptability likelihood of the strategies. The following paragraphs discuss the behavioral strategies and the corresponding modeling framework adopted to represent both the propensity toward each option and the resulting aggregate selection patterns.

\paragraph{Probability of rigidity and time-shifting:}

The formulation of the model for rigidity and time-shift is already present in the literature, and it is known as the Exponentiated Linear Elasticity Model \citep{Hursh1988CostBenefit, Koffarnus2022Behavioral}. Our formulation is a special case of the model, where the slope and coefficient of the demand curve are set to $0$.

Let $X$ denote the variable of maximum acceptable effort. Let us assume that this variable follows an exponential distribution of density $f(x) = \lambda \exp\{-\lambda x\}$. Among the behavioral parameters of the model, it is possible to specify the median of this distribution, $X_{50\%}$, that corresponds to the acceptable effort for 50\% of travelers, and that is related to the parameter $\lambda$ through the formula $\lambda = \ln (2) / X_{50\%}$. 

\noindent Then, once the effort required by the access policy $X_0$ is known, it is possible to compute the probability that the required effort is less than the acceptable effort, as in Equation \ref{eq:prob}. This probability also represents the acceptance rate of the strategy, thereby defining the share of travelers allocated to that behavioral response. 

\begin{equation}\label{eq:prob}
    \mathbb{P}[\text{accept the strategy}] = \mathbb{P}[X\geq X_0] = \exp\left\{- \frac{\ln(2)}{X_{50\%}} \cdot X_0\right\}.
\end{equation}

Equation \ref{eq:prob} is applied to estimate the probability of accepting rigidity. In this context, the effort corresponds to the monetary cost of the policy per emission level $l$, $C_{0l}$, compared with the maximum acceptable cost, $C$, exponentially distributed with median $C_{50\%}$. Using Equation \ref{eq:prob:rigid}, it is possible to calculate the probability of accepting rigidity per emission level as $\mathbb{P}[C\geq C_{0l}]$ during the policy application:

\begin{equation}\label{eq:prob:rigid}
    \mathbb{P}[\text{rigidity for Euro-level } l] 
    = \mathbb{P}[C\geq C_{0l}] 
    = \exp\left\{- \frac{\ln(2)}{C_{50\%}} \cdot C_{0l}\right\}.
\end{equation}

\noindent By reweighting the probability with the proportion of vehicles belonging to each emission level class, the probability of overall rigidity acceptance is found, as in Equation \ref{eq:prob:rigid2}:

\begin{equation}\label{eq:prob:rigid2}
    \mathbb{P}[\text{rigidity}] 
    = \sum_l P_l \cdot \mathbb{P}[\text{rigidity for Euro-level } l] 
    = \sum_l P_l \cdot \mathbb{P}[C\geq C_{0l}].
\end{equation}

The postponement of trips is evaluated by assessing the effort of a time variation from the current travel starting time to the policy ending time, $\Delta_{te}(t) = t_e-t$ $\forall t\in[t_s,t_e]$. The time variation is to be compared with the maximum acceptable time variation, $\Delta_p$, exponentially distributed with median $\Delta_{p50\%}$. As $\Delta_{te}$ depends on the time, the probability of postponement is also time-dependent on each starting time interval $t$ throughout Equation \ref{eq:prob:postponement}:

\begin{equation}\label{eq:prob:postponement}
     \mathbb{P}[\text{postponement in time } t] 
    = \mathbb{P}[\Delta_p \geq \Delta_{te}(t)] 
    = \exp\left\{- \frac{\ln(2)}{\Delta_{p50\%}} \cdot \Delta_{te}(t)\right\}.
\end{equation}

\noindent Similarly, Equation \ref{eq:prob:anticipation} uses the acceptance anticipation $\Delta_a$, the required anticipation per time interval $\Delta_{fs}(t)$, and the median parameter $\Delta_{a50\%}$, to compute the probability of accepting anticipation. However, to fully avoid the circulation policy, the trip should be anticipated to be completed before the policy starts. Consider the travel time in the area, modeled as a random variable with a geometric distribution $\tau\sim \mathcal{G}eom(p=\tfrac{1}{\bar{\tau}})$ over the support $\mathbb{N}_{>0}$. Then the probability of anticipation can be found as in Equation \ref{eq:prob:anticipation}:

\begin{equation}
\begin{split}\label{eq:prob:anticipation} 
    \mathbb{P}[\text{anticipation in time } t] 
    = & \sum_{T=1}^{\infty} \mathbb{P}[\Delta_a \geq \Delta_{fs}(t) + \tau | \tau = T ] \cdot \mathbb{P}[\tau=T]  \\
    = &\: \varepsilon \cdot \exp\left\{- \frac{\ln(2)}{\Delta_{a50\%}} \cdot \Delta_{fs}(t)\right\},\\
\end{split}
\end{equation}
where $\varepsilon = p\, \exp\{-{\ln(2)}/{\Delta_{a50\%}}\} [1-(1-p) \exp\{-{\ln(2)}/{\Delta_{a50\%}}\}]^{-1}$.

\paragraph{Probability of mode-shifting:} 

The probability of the change in the mode of transportation is modeled through a logit model \citep{BenAkiva1999Discrete}. It is based on the principle that the probability of changing between two modes of transport depends on the relative effort of the options. In this case, since focusing on the shift from private to public transport, the model incorporates covariates related to the availability of the public transport system (PTS), as well as indicators of the overall convenience of making the change, to assess the modal shift proportion. 
As the public transport distribution is not homogeneous in the city, we characterize subzones of the metropolitan areas with their indicators of availability and convenience, and apply independent modal shift formulations for each travel zone. Consequently, each external origin zone generating trips to the area is assigned a specific probability of modal shift.

Among the considered covariates, two of them represent the objective characteristics of public transport supply and their adequacy relative to travelers' needs per zone: $x_{1z}$ is the frequency and $x_{2z}$ is the capillarity in zone $z$.
The frequency is defined as the passage frequency of public transport at stops, and is considered an index of service availability. For each stop and line passing through it, the average number of hourly passages is calculated. By then aggregating this value by zone, the average passage frequency of the vehicles at the stops in the area is obtained.
The capillarity is defined as the proportion of the transportation network covered by the PTS, and is considered a proxy of the service accessibility. It is calculated per zone as the ratio of the transport network length served by public transit routes, determined by matching the network to the public transit routes, to the total transport network length.

Two additional covariates, then, evaluate the convenience of public transport compared to private transport: $x_{3z}$ is the cost and $x_{4z}$ is the time difference in the zone $z$.
The cost of the public transportation service is the fare that a user would pay to travel via public transportation from any zone to the area of interest, evaluating the economic convenience of the mode-shift. The difference in travel times when using public transport and when using private means, then, assesses the temporal convenience of the mode-shift. 

Let $p_z$ be the probability of accepting a mode shift to the public transport when starting the trip in zone $z$. Then, its logit is related to the linear combination of the transport covariates, $x_{1z}, \dots,x_{4z}$., as in the following formulation:
\begin{equation} \label{eq:prob:mode}
    \text{logit}(p_z) = \ln \frac{p_z}{1-p_z} = \beta_0 + \sum_{i=1}^4 \beta_i x_{iz}.
\end{equation}
\noindent The coefficients $\beta_0, \dots, \beta_4$ reflect the relative importance assigned to each attribute in the modal choice by travelers. They are not estimated through regression but are treated as behavioral parameters, set in advance.

For each zone of origin, the probability of the modal shift is given by Equation \ref{eq:prob:mode2}:
\begin{equation}\label{eq:prob:mode2}
    \mathbb{P}[\text{modal shift from zone }z] = \text{logit}^{-1}\left(\beta_0 + \sum_{i=1}^4 \beta_i x_{iz}\right).
\end{equation}

\noindent By reweighting the probability with the proportion of vehicles starting in each zone, $Z_z$, the probability of overall modal shift acceptance is found as in the following Equation:
\begin{equation}\label{eq:prob:mode3}
    \mathbb{P}[\text{modal shift}] = \sum_z Z_z \cdot \mathbb{P}[\text{modal shift from zone }z].
\end{equation}

\paragraph{Aggregate strategy-choice pattern:}
 
After defining the set of available behavioral options, the model represents how these alternatives are comparatively evaluated at the aggregate level. Rather than describing individual cognitive processes, the framework specifies alternative evaluation rules that approximate collective behavioral patterns observed in similar contexts. For example, options may be assessed either sequentially (by assigning priority to certain behaviors) or simultaneously (by comparing all alternatives within a common evaluative structure). In this study, a simultaneous and independent evaluation of options is adopted as a modeling assumption, and the selected strategy corresponds to the alternative associated with the lowest perceived effort at the aggregate level. 
This represents a modeling choice; however, the framework can readily accommodate alternative models, such as the sequential evaluation strategy or hybrid approaches that combine multiple decision rules.

Equations \ref{eq:prob:rigid2}, \ref{eq:prob:anticipation}, \ref{eq:prob:postponement}, and \ref{eq:prob:mode3} report the marginal probabilities that a specific strategy is acceptable within the modeling framework. These probabilities are specified independently, such that the acceptability of one strategy does not affect the acceptability of another.
As such, the framework allows for multiple possible configurations: a single strategy may be acceptable, none may be acceptable, or several strategies may simultaneously satisfy the acceptability condition. In the case of acceptance of multiple strategies, the model requires an allocation rule to determine the implemented outcome. 
%
Specifically, each acceptable strategy is assigned a weight that determines its probability of being selected. These weights can be uniform, reflecting a situation in which all acceptable strategies are equally likely to be chosen (e.g., analogous to a random selection), or they can reflect relative probabilities that capture preferences, likelihoods, or other decision criteria. 

Note that, since the marginal probabilities of anticipation and postponement are time-dependent, also the fraction of rigid varies with time $t\in[t_s,t_e]$. 
In general, given any strategy $X$ chosen from those available, $E=\{\text{rigidity}, \text{anticipation}, \text{postponement}, \text{mode-shift}\}$, the fraction of travelers with such a strategy is calculated in Equation \ref{eq:prob:final:general}, obtaining respectively $F_{rig}, F_{ant}, F_{post},$ and $F_{mode}$:

\begin{equation}\label{eq:prob:final:general}
    F_X(t) = (1-F_e) \cdot \sum_{S\subseteq E; X\in S} w(X, S,t)\prod_{Y\in S} \mathbb{P}[Y](t) \prod_{Z\notin S}(1-\mathbb{P}[Z](t)),
\end{equation}

\noindent where $S\subseteq E$ is any combination of strategies including $X$; and $w(X, S, t)$ represents the weight associated to the acceptability of strategy $X$ within the set $S$ at time $t$. In the current implementation, we use the uniform weights: $w(X,S,t)=1/|S|$. However, as discussed, alternative weighting schemes can be used within the same framework, such as weights proportional to the marginal probabilities: $w(X,S,t)={\mathbb{P}[X](t)}/{\sum_S \mathbb{P}[Y](t)}$.

Those for which none of the strategies are acceptable by the model become part of the \textit{lost} vehicles, either canceling the trip, or changing to other transport modes rather than public transport, or choosing other options not accounted for in the model. The fraction of lost vehicles is computed as the fraction of non-exempted vehicles, multiplied by the probability that a vehicle falls into the category of non-acceptable strategies, as in Equation \ref{eq:prob:final:nothing}.

\begin{equation}\label{eq:prob:final:nothing}
\begin{split}    
    F_{lost}(t) = (1-F_e) \cdot & 
    \Big((1-\mathbb{P}[\text{rigidity}]) 
    (1-\mathbb{P}[\text{anticipation}](t)) \\ 
     &
    (1-\mathbb{P}[\text{postponement}](t)) 
    (1-\mathbb{P}[\text{modal-shift}])\Big). \\
    \end{split}
\end{equation}

\paragraph{Strategies for time-shifting:} Having determined the proportion of vehicles adopting each behavioral strategy, most categories can be directly processed: rigid and exempt vehicles contribute to the calculation of the inflow during policy time slots, while mode-shifted and lost vehicles are removed from the analysis. On the other hand, time-shifting vehicles require an additional layer of analysis. 
Besides representing whether vehicles shift their departure time, the model also specifies the magnitude of the anticipation or postponement. For example, for vehicles assigned to an earlier departure strategy, the model determines the corresponding timing adjustment according to a prescribed allocation rule. This temporal reallocation introduces another modeling phase, as these vehicles must remain in the daily total count while being redistributed across different time intervals according to specific shifting behaviors.


This model accounts for a time-shifting approach that separates the shifting decision into two stages. First, vehicles are classified according to whether the time-shifting strategy applies, based on the model's criterion of avoiding the policy period, regardless of specific alternative timings. Vehicles assigned to this strategy are then redistributed to new departure times before or after the policy period, independently of their original departure times.

Specifically, using Equation \ref{eq:prob:final:general} from the first phase, the fraction and, therefore, the number of vehicles assigned to a time shift are identified. Then, the total of vehicles anticipating $TN_{ant}$ and postponing $TN_{post}$ are calculated as in Equation \ref{eq:total:ap}.
\begin{equation}\label{eq:total:ap} \begin{split}
    TN_{ant} & = \sum_{t\in[t_s, t_e]} F_{ant}(t) \cdot I(t), \\
    TN_{post} & = \sum_{t\in[t_s, t_e]} F_{post}(t) \cdot I(t). \\
\end{split}\end{equation}

Through the second phase, the redistribution of these totals is modeled both before and after the policy's duration. The reasoning applies to both anticipation and postponement, but let us now consider the case of postponement. 
The random variable $\Delta_{pr}$ indicates the time variation associated with the postponement strategy, specifically, the interval between the assigned new departure time and the end of the policy period.
The exponential distribution is chosen for $\Delta_{pr}$, 
with median $\Delta_{pr50\%}$, scale parameter equal to $\lambda=\ln (2)/\Delta_{pr50\%}$, and the density $f_{pr}(x) = \lambda \exp\{- \lambda x\}$.
When determining how many trips will be moved to a specific time interval $t$, the analysis requires the density corresponding to the specified timing adjustment, representing the fraction of vehicles assigned by the model to a shift from the policy end of exactly the magnitude $\Delta_{fe}(t)$.
This density directly gives the rate at which trips concentrate at time interval $t$, computed as in Equation \ref{eq:antic:1}. 
The same reasoning applies symmetrically to anticipations, where we consider the random variable $\Delta_{ar}$  with median $\Delta_{ar50\%}$, density $f_{ar}$, and the time variation is measured until the start of the policy period, denoted as $\Delta_{ts}(t)$.

\begin{equation}\label{eq:antic:1}\begin{split}
    F_{pr}(t) = & \mathbb{P}[\Delta_{fe}(t-1)\leq \Delta_{pr} \leq \Delta_{fe}(t)] =  \exp\left\{-\frac{\ln(2)}{\Delta_{pr50\%}} \Delta_{fe}(t-1)\right\} - \exp\left\{-\frac{\ln(2)}{\Delta_{pr50\%}} \Delta_{fe}(t)\right\}, \\
    F_{ar}(t) = & \mathbb{P}[\Delta_{ts}(t-1)\leq \Delta_{ar} \leq \Delta_{ts}(t)] =  \exp\left\{-\frac{\ln(2)}{\Delta_{ar50\%}} \Delta_{ts}(t-1)\right\} - \exp\left\{-\frac{\ln(2)}{\Delta_{ar50\%}} \Delta_{ts}(t)\right\}. \\
\end{split}\end{equation}
\noindent The absolute number of anticipated redistributed vehicles at each time interval is then computed as the fraction $F_{ar}(t)$ multiplied by the total anticipating $TN_{ant}$. Similarly, the number of postponed redistributed vehicles is assessed by multiplying $F_{pr}(t)$ by $TN_{post}$.
\begin{equation} \begin{split}
    N_{ar}(t) = & F_{ar}(t) \cdot TN_{ant}, \\
    N_{pr}(t) = & F_{pr}(t) \cdot TN_{post}. \\
\end{split} \end{equation}

\paragraph{Inflow in the modified scenario: } Given the proportion of vehicles assigned to each behavioral strategy, it is possible to calculate the number of vehicles entering the policy-affected area under the modified scenario. Equation \ref{eq:inflow_mod} presents the formulation for computing the inflow, which distinguishes between two cases depending on policy application time. Within the period of enactment of the policy, with $t\in[t_s,t_e]$, the inflow is the sum of exempted vehicles and rigid vehicles. Outside the hours of the policy, with $t\notin[t_s,t_e]$, the inflow is modified from a portion of vehicles that anticipate and postpone the starting time and sum to the original vehicles.
\begin{equation}\label{eq:inflow_mod}
    I_m(t) = 
    \begin{cases}
        \Big(F_e + F_{rig}(t) \Big)\cdot I(t) & t\in[t_s,t_e], \\
       I(t) + N_{ar}(t) + N_{pr}(t) & \text{otherwise}.\\
    \end{cases}
\end{equation}

\subsection{Traffic estimation model} \label{subsec:traffic}

In addition to the entrances, the policy affects the amount of traffic in the area. Our methodology does not attempt to model traffic variations directly; rather, its purpose is to capture the dependence of traffic on selected quantities, specifically inflows and startings, allowing the estimation of modified traffic conditions when these quantities change.

The total traffic load is generated by two sources: trips entering from outside the area (i.e., inflows) and trips starting within the area itself (i.e., starting trips). At any time interval $t$, one component of the traffic load consists of newly initiated trips, both inflow and starting trips, that begin during that interval.
However, traffic does not reset at each time interval but evolves progressively as trips progress through the network. 
Many trips that are active at time $t-1$ remain active at time $t$, creating a temporal dependency between consecutive observations. This persistence means that the current traffic state is not independent from the previous one, but rather builds upon it. Consequently, the remaining traffic component is given by a fraction of the previous traffic load that persists into the current time step. Quantifying this persistence through a retention probability allows us to relate the evolution of traffic load to the pattern of trip departures over time.

Another key feature of traffic is its daily periodicity with characteristic recurrent patterns of morning and afternoon peaks and night-time lows. To formalize this periodicity, we define the set of observation times $\mathcal{T} = \{t_0, t_1, \dots, t_{n-1}\}$ as a discrete time space with equivalence modulo of 24 hours. This means that each integer time index $s \in \mathbb{N}$ can be associated with a representative $t \in \mathcal{T}$ such that
$t = s \bmod n$, where $n$ is the number of observation intervals in a day. This construction ensures that operations like $t-1$ are always well-defined: for example, the interval immediately before the first observation after midnight corresponds naturally to the last interval of the previous day. In other words, the day is treated as a looped discrete-time system, simplifying the treatment of temporal dependencies at the boundaries of the day.

Given the inflows and starting trips, we model traffic as a looped discrete-time system with observation intervals of $\nu$ (e.g., 5 minutes). 
At any time $t \in \mathcal{T}$, $I(t)$ represents vehicles entering from outside (i.e., the inflow) and $S(t)$ represents vehicles starting their journey within the area.
Then, traffic evolves according to the following Equation \ref{eq:traffic_evol}:
\begin{equation}\label{eq:traffic_evol}
T(t) = I(t) + S(t) + \alpha \cdot T(t-1) \quad \forall t \in \mathcal{T},
\end{equation}
where $\alpha$ is the retention probability, representing the fraction of vehicles circulating at time $t-1$ that will remain in the area at time $t$. 
This formulation combines both the {incremental contribution of new trips} and the {persistence of existing traffic}, while the looped time space $\mathcal{T}$ guarantees consistency across the limits of the day.

Our approach models the retention probability, $\alpha$, based on assumptions about the dwell time in the area, i.e., the duration for which a vehicle remains active within the area. 
We make two key assumptions about vehicle dwell times. First, the probability that a vehicle departs at a given moment does not depend on how long it has already been waiting. Second, dwell times are assumed to be independent of traffic conditions, so the probability of departure is stable regardless of congestion. Given this modelling choices, we consider the dwell times to be geometrically distributed with constant mean \citep{Grimmett2001Geometric}. 
Within this framework, the retention probability $\alpha$ represents the probability that a trip does not end in a given time step and can be directly related to the mean dwell time. 

Specifically, let $\tau$ be the trip dwell time following a geometric distribution with mean dwell time $\bar{\tau}$, that is, $\tau\sim \mathcal{G}eom(p=\tfrac{1}{\bar{\tau}})$ over the support $\mathbb{N}_{>0}$.
This yields to the formulation of the retention probability $\alpha$ as a function of the average dwell time $\bar{\tau}$, as $\alpha = \frac{\bar{\tau} - 1}{\bar{\tau}}$.

The objective is to compute the traffic vector satisfying Equation \ref{eq:traffic_evol}. 
We adopt an iterative approach as outlined in Algorithm \ref{alg:traffic-estimation}.
The starting point is an initial guess for the traffic values, $T(t)^0 = I(t) + S(t)$ for all $t\in\mathcal{T}$. Then, at each iteration step $k$, the algorithm recomputes updates the traffic estimate $T^{k+1}$, as shown in Equation \ref{eq:traffic_update}:
\begin{equation}\label{eq:traffic_update}
        T(t)^{k+1} = I(t) + S(t) + \alpha \cdot T(t-1)^k \quad \forall t\in\mathcal{T} \quad \forall k>0.
\end{equation}
The estimation is repeated until the traffic vector $T$ satisfies the initial Equation \ref{eq:traffic_evol} given a certain tolerance $\gamma$. Convergence is ensured by the fact that $\alpha < 1$.

\begin{algorithm}[tb]
\footnotesize
\caption{Traffic Estimation Algorithm.} 
\label{alg:traffic-estimation}

\KwIn{
\\
- Set of observation interval $\mathcal{T}=\{t_0,\dots,t_{n-1}\}$
- Inflow and starting vectors $I, S \in\mathbb{R}_{\geq 0}^n$\\
- Average reference dwell time $\bar{\tau}\in \mathbb{N}_{>0}$ (expressed in $\nu$-min intervals)\\
- Convergence threshold $\gamma \in \mathbb{R}_{>0}$
}
\BlankLine\BlankLine
$\alpha \leftarrow \frac{\bar{\tau} - 1}{\bar{\tau}}$ \tcc*{Retention probability} 

$T^{(0)}(t) \leftarrow I(t) + S(t) \quad \forall t \in \mathcal{T}$ \tcc*{Initialize traffic vector} 

\While{$\max_t |T^{(k)}(t) - T^{(k-1)}(t)| > \gamma$}{

    $T^{(k+1)}(t) \leftarrow I(t) + S(t) + \alpha \cdot T^{(k)}(t-1) \quad \forall t \in \mathcal{T}$ \tcc*{Update traffic vector} 
}
\BlankLine\BlankLine
\KwOut{$T^{(k)}$ \tcc*{Estimated traffic profile}}
\normalsize
\end{algorithm}

Having described the traffic model, we now address the computation of its input data, which are required to estimate the traffic vectors in both the reference and policy scenarios. 
First, in the reference scenario, $T$ is computed using the inflow and starting vectors $I$ and $S$. The inflows $I$ and the starting trips $S$ are extracted from observed data representing the current situation.
Then, in the policy scenario, the modified traffic $T_m$ is estimated starting from the corresponding modified vectors $I_m$ and $S_m$, where $I_m$ is adjusted according to the procedures outlined in the previous Subsection~\ref{subsec:inflow}. 
For the starting vector, the same reasoning applied to the inflow can be adopted. All non-exempt vehicles starting within the area of interest are subject to the ticket fee imposed by the policy, and their behavioral response is represented through the four adjustment strategies of rigidity, time-shifting, mode-shifting, or trip cancellation. The modified starting vector is computed following the same procedure used for the inflow case, replacing the inflow vector $I$ with the starting vector $S$.

\subsection{Emission estimation model} \label{subsec:emissions}

For the calculation of emissions in the area, it is assumed that each vehicle in circulation produces a value of average emissions per kilometer depending on its characteristics. This work uses the classification of vehicles in Euro emission standards \citep{Selleri2021Overwiew}, widely adopted in European environmental and transport studies, to standardize the assessment of vehicle emissions and ensure comparability across different regions and datasets. Vehicles are classified into seven Euro classes, from Euro-0 to Euro-6, based on their characteristics such as vehicle type, fuel type, date of the first registration, and maximum emissions per kilometer. Accordingly, we define $\bar{E}_{l, km}$ as the representative average emissions per kilometer corresponding to the $l$-th Euro-class, with values obtained from the literature. Moreover, let $P_l$ represent the percentage of vehicles for each of the Euro-class $l$. This proportion reflects the actual composition of the fleet in the current base scenario, derived from the elaboration of real-world data. 

The distribution of vehicles varies before and after the implementation of the policy, as the policy imposes different costs on different Euro-classes, which in turn affects the likelihood of vehicles being rigid. Consequently, during the period of application of the policy, the Euro-level distribution of the fleet must be recalculated considering the adjusted proportions of rigid vehicles, while outside this period it must be based on the original distribution.
As a first step, the rigidities for each Euro-class must be calculated. Equation \ref{eq:frac:rigid:el}
represents the allocation of the total rigidity fraction, $F_{rig}$, among different Euro-levels $l$. In this way, the formula distributes the total rigidity among the partitions proportionally to both the size of the partition and the likelihood of rigidity within it for each time interval $t\in[t_s, t_e]$.
\begin{equation} \label{eq:frac:rigid:el}
    F_{rig, l}(t) = F_{rig}(t) \cdot  \frac{\mathbb{P}[\text{rigidity for Euro-level }l] }{\sum_j P_j \cdot \mathbb{P}[\text{rigidity for Euro-level }j] }.
\end{equation}
Then, it is possible to compute the Euro-level distribution under the policy enactment. During the period $[t_s, t_e]$, the new proportion $P_{l, m} (t) $ accounts for two contributions. First, it considers the fraction of exempt vehicles $P_l\cdot F_e$, which represents the part of the Euro-class $l$ that is not subject to the policy and remains unchanged. Second, it considers the fraction of rigid vehicles, $P_l \cdot \frac{F_{rig,l}(t)}{F_e + F_{rig}(t)} $, which represents the proportional contribution of the rigid share of class $l$ compared to the total of rigid vehicles and exempt vehicles. Equation \ref{eq:modified-euro-split} formalizes this reasoning.
\begin{equation} \label{eq:modified-euro-split}
    P_{l,m}(t) = 
    \begin{cases}
        P_l \cdot \left( F_e + \frac{F_{rig,l}(t)}{F_e + F_{rig}(t)}\right) & t\in [t_s, t_e], \\
        P_l & \text{otherwise.}\\
    \end{cases}
\end{equation}

Finally, the average emission value per vehicle in the time interval $t$ is calculated as in Equation \ref{eq:emission_1}. The formula is applied for the computation of emissions in both scenarios. 
In this context, the variable $km\nu$ serves as a conversion factor between distance traveled and the corresponding temporal interval elapsed $\nu$. Specifically, it is assumed that each vehicle, regardless of its Euro-level, travels an average distance of $km\nu$ for each unit of time $\nu$.
\begin{equation}\label{eq:emission_1}
\begin{split}
    \bar{E}(t) &= \sum_l \left( P_l(t) \cdot km\nu \cdot \bar{E}_{l,km} \right), \\
    \bar{E}_m(t) &= \sum_l \left( P_{l,m}(t) \cdot km\nu \cdot \bar{E}_{l,km} \right). \\
\end{split}
\end{equation}

The total emissions produced can be determined by multiplying the number of vehicles $T(t)$ in circulation during each time period $t$ by the average emissions per vehicle, as in Equation \ref{eq:emissions_2}. The formula is applied for the computation of emissions in both scenarios.
\begin{equation} \label{eq:emissions_2}
\begin{split}
    E(t) &=  \bar{E}(t) \cdot T(t), \\
    E_m(t) &= \bar{E}_m(t) \cdot T_m(t).\\
\end{split}\end{equation}

\section{Implementation and case study}\label{sec:application}

\subsection{Technical development} \label{subsec:tech}

The implementation of the proposed system requires work on two complementary fronts: the backend and the front end. On the backend side, the focus is on ensuring efficient computation and robust model execution, handling large-scale mobility and traffic data, and implementing the algorithms described in the previous sections. On the front end, the objective is to provide an accessible and intuitive interface that could be used by domain experts with varying levels of technical expertise. Together, these two components form an integrated technological system that balances computational rigor and usability, and enables both precise estimation of traffic outcomes and interactive policy experimentation.

The backend is responsible for model development and computation. Most implementation is collected in the Python package \texttt{civic-digital-twin}, which can be found in the relative GitHub page \citep{github2025cdt}. 
As previously mentioned, every variable in the model is represented as an index. Independent indices can be assigned either fixed point values or parametric distributions, from which values are sampled during the different iterations. Each index is characterized by a name and, if independent, a value; for dependent indices, the relationships to other variables are defined, specifying both the dependencies and their analytical form. This structure enables the construction of a dependency graph, allowing the sequential computation of dependent variable values at each iteration.
The entire computation framework is based on an ensemble modeling approach, in which independent models are evaluated in parallel and their outputs aggregated. This approach has a twofold advantage: it reduces the variance of prediction errors and allows the uncertainty of the results to be quantified, for example, in the form of confidence intervals. 

The frontend is designed to ensure accessibility and usability for domain experts, including those without technical backgrounds. To this end, we develop an interactive dashboard that allows near real-time simulation of what-if scenarios under varying conditions. Users can directly modify key control parameters, described in Sections \ref{sec:meth} and \ref{sec:math}, including policy fees per vehicle type, exempted vehicles, temporal windows of the policy application, behavioral parameters, and public transport settings. 
On the same page, simulation results are presented clearly and intuitively (see Section \ref{sec:app:results} for a wider discussion on qualitative and quantitative results). Global indicators summarize daily information for both the as-is baseline and  what-if modified scenarios. They refer to indicators such as the total vehicle inflow per day, the maximum traffic within the day, total collected fees, and the estimated emission modification. For insight at a microscopic scale, interactive graphs are included to display the daily evolution of key quantities of interest in the current and modified scenarios.
By changing the parameters, users can evaluate multiple scenarios corresponding to different combinations of parameters. The dashboard not only enables comparison between the baseline and modified scenarios, but also across different what-if scenarios (e.g., scenario 1 vs. scenario 2), supporting the assessment of alternative policy options.

\subsection{Application: the case of Bologna, Italy}

\begin{figure}[tb]
    \centering
    \includegraphics[width=0.27\textwidth]{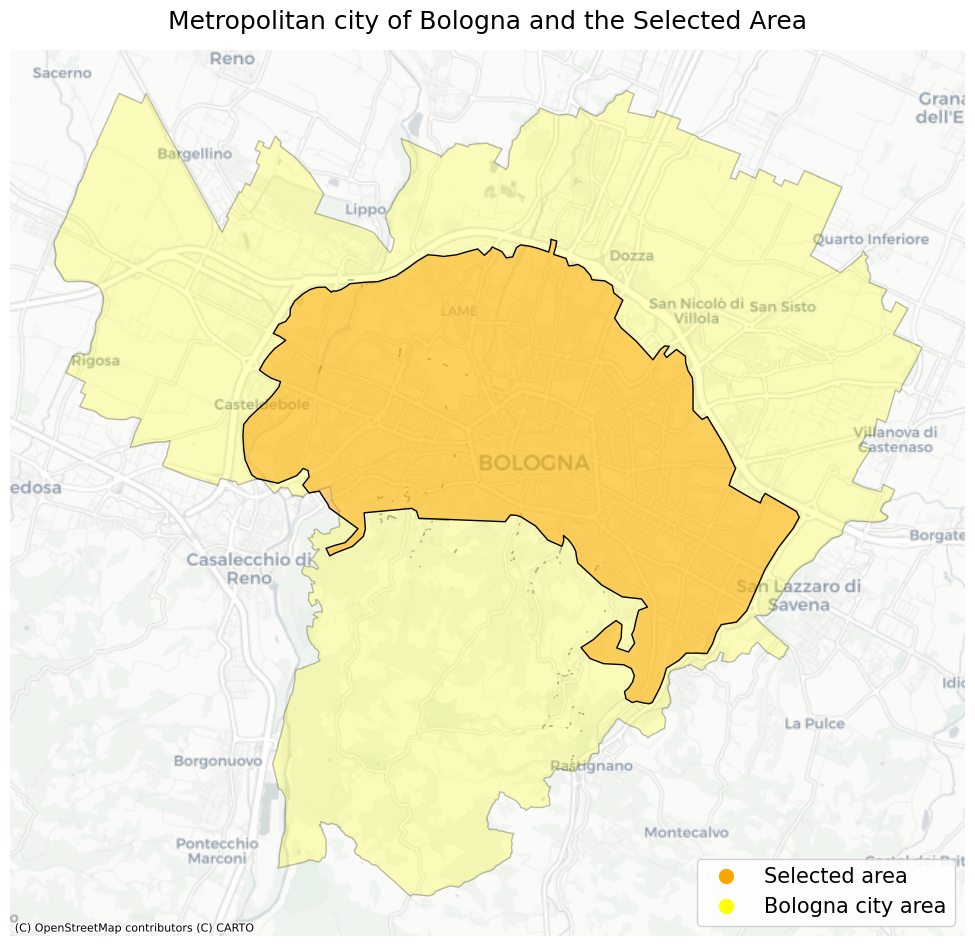}\hfill
    \includegraphics[width=0.32\textwidth]{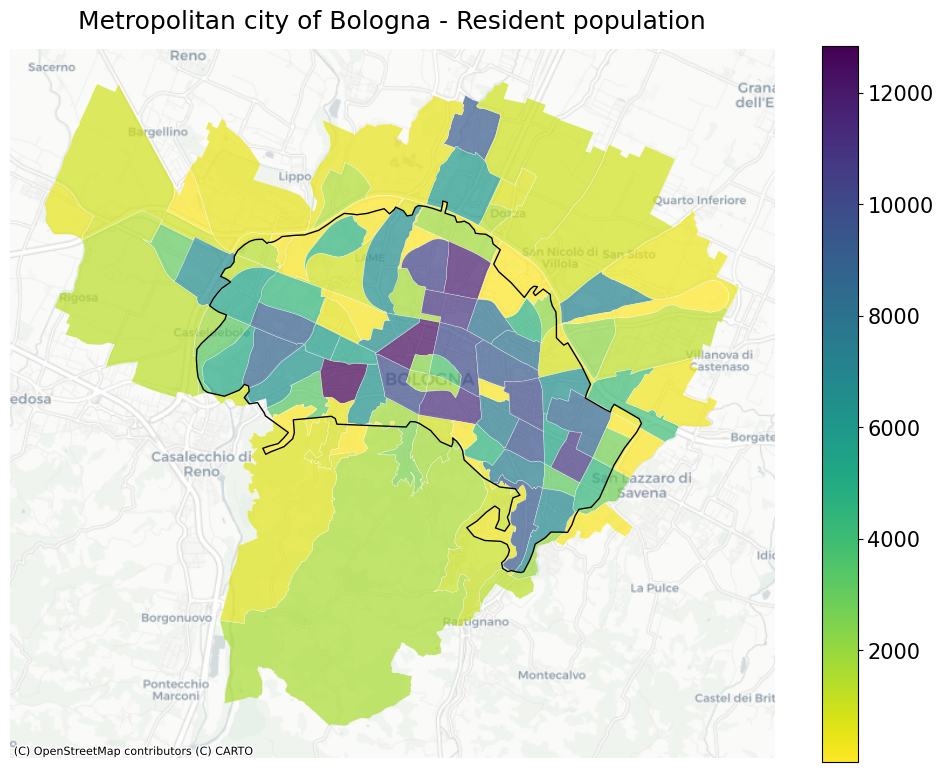}\hfill
    \includegraphics[width=0.27\textwidth]{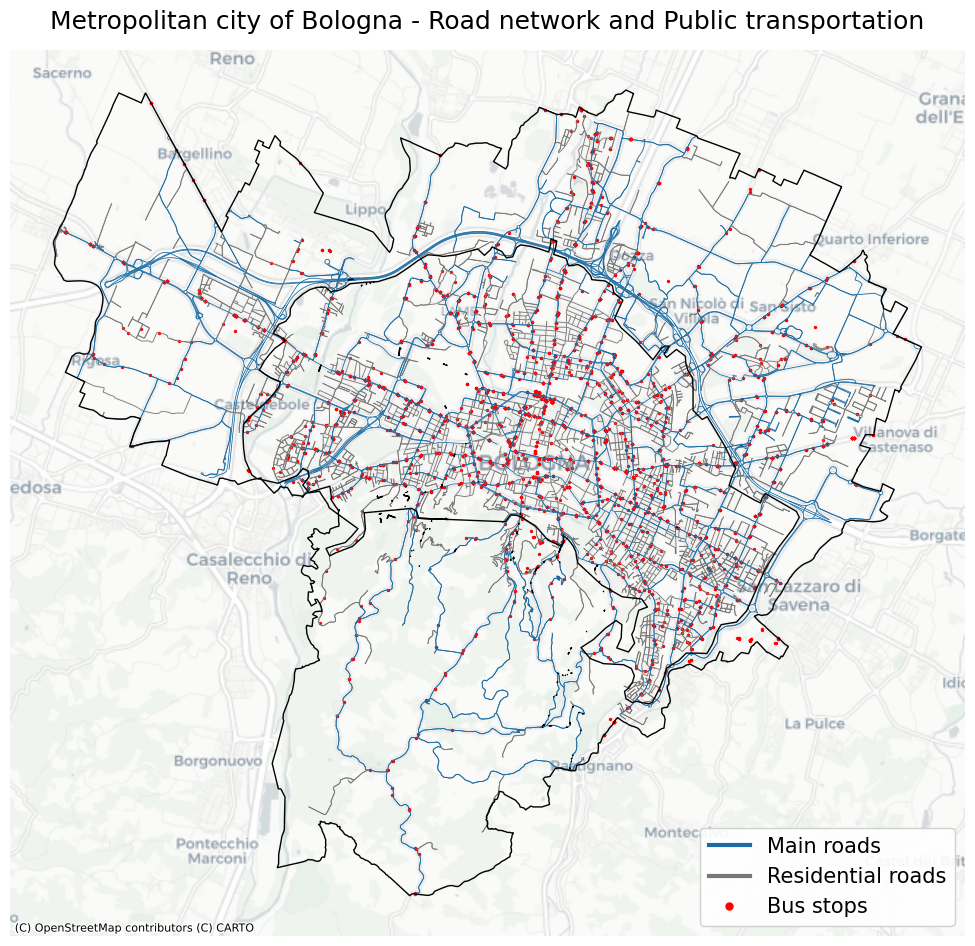}
    \caption{Left: the city of Bologna is highlighted in yellow, while the selected area, subject to the policy, is colored in orange. Center: population distribution in Bologna; the boundary of the selected area is highlighted in black. Right: road network and public transport stops in Bologna; the boundary of the selected area is highlighted in black.}
    \label{fig:area}
\end{figure}

The proposed model is tested on a real-world case study in the city of Bologna in Italy, a major urban center with a large resident population and a complex multimodal mobility system. The city is already characterized by a limited traffic zone in the historical center. For experimental purposes, we simulate an additional circulation restriction applied within a broader zone: its northern boundary is delineated by the ring road and the A1-A14 motorway, while the southern boundary follows the city’s morphological constraints, namely, the hilly area. Figure \ref{fig:area} represents the selected area. This configuration serves as an illustrative example of how the model can be applied to large-scale area-based policies.

The selected area exhibits both strong centrality and high urban attractiveness. It includes more than $321,172$ residents, $82$\% of Bologna’s $392,044$ urban population\footnote{Data extracted from the “Resident Population by Age, Sex, Citizenship, Neighborhood, and Statistical Area” dataset from the Comune di Bologna Open Data portal, version updated 31/12/2024; accessed 12/12/2025. URL: \url{https://opendata.comune.bologna.it}.} and $31$\% of the $1,051,536$ residents in the wider metropolitan province\footnote{Data extracted from the “Resident Population” ISTAT dataset; accessed on 12/12/2025. URL: \url{https://demo.istat.it}.}. It concentrates substantial private motorized mobility flows, with $269,226$ daily trips entering the area from outside and $329,807$ internal trips\footnote{Origin-Destination data are extracted and relaborated from PUMS 2019, provided by Comune di Bologna.}. The road network of the selected area covers approximately $708$ km of drivable roads, of which $287$ km correspond to primary corridors (i.e., with tags trunk, primary, secondary, and tertiary roads according to OpenStreetMap)\footnote{Road network data extracted from OpenStreetMap; accessed 8/06/2025. URL: \url{https://www.openstreetmap.org}.}. Public transport is also widely available in the area. In particular, for weekdays, the bus routes cover $249$ km of the drivable road network with $177$ bus lines, operating on a total of $786$ bus stops with a median frequency of $14$ trips per line per day (range: $1$ to more than $100$)\footnote{Bus service data extracted from the Open Data portal of Trasporto Passeggeri Emilia‑Romagna (TPER); accessed on 25/05/2025. URL: \url{https://solweb.tper.it/web/tools/open-data/open-data.aspx}.}. The area also includes Bologna Centrale, a major national and regional railway hub. The station serves high-speed and intercity trains to major Italian cities (e.g., Florence, Milan, Venice, Rome) as well as international connections, including routes to Munich, and regional and suburban lines connecting surrounding towns and the Modena-Bologna-Ferrara corridor.

Bologna is further selected due to the advanced mobility data infrastructure maintained by the municipality. High-resolution data are collected from multiple sources, including traffic sensors, cameras, and vehicle registration data from the national motorization authority                                                                     . These datasets allow us to estimate the current as-is traffic situation. In particular, the integration of multi-source data with statistical and data science preprocessing techniques enables the estimation of inflows into the restricted area as well as trips originating within it, with a temporal discretization of $\nu=5$ minutes. Furthermore, we have estimates of the daily percentage of vehicles by Euro emission class currently circulating in the area. From Euro-0 to Euro-6 vehicles, respectively, the proportions are $0.059$, $0.012$, $0.034$, $0.054$, $0.198$, $0.176$, and $0.467$. Similarly, knowing the vehicular fleet composition in terms of segment, fuel type, and technology, we estimate the pollutant emissions per kilometer per vehicle typology in the area. We focus on the nitrogen oxides (NOx), a group of highly reactive gases emitted by vehicles that significantly contribute to air pollution. Although NOx is our primary subject, this analytical framework can be equally applied to other components of vehicular air pollution (e.g., particulate matter [PMx], carbon monoxide [CO], carbon dioxide [CO2]). For each Euro-level class, the estimated emissions in grams per kilometer are: $0.2105847$, $0.217457$, $0.2401457$, $0.247239$, $0.135555$, $0.099559$, and $0.068246$.


\subsection{Result assessment}\label{sec:app:results}

In this section, we present the outcomes of the simulation scenarios under different access pricing policies and behavioral settings. To explore the potential effects of alternative parameter configurations, we construct a set of scenarios and systematically compare their impacts. The purpose of this analysis is not primarily to evaluate specific policies, but rather to illustrate a methodological approach for measuring variations in the model outputs under different conditions with the proposed methodology. The focus is on demonstrating how scenario-based simulations can be used to quantify and analyze changes in key performance indicators.

\subsubsection{Scenario definition}\label{subsub:scenario}

In addition to the baseline scenario, we consider a set of alternative scenarios designed to illustrate the effects of different pricing configurations and target populations. These scenarios are selected to represent not only typical, realistic situations but also extreme cases that test the limit of the model. By including both ends of the spectrum, we can explore how the system responds under a variety of conditions and identify which parameters have the greatest influence on outcomes. This approach allows a systematic comparison across scenarios, providing insights into the sensitivity of the case study to the choice of parameter settings.

First, we propose a scenario analysis aimed at highlighting the effects of different policy timing schemes. We examine a daily fixed-price policy applied to the entire population during a 10h period from 8:00 AM to 6:00 PM (scenario A1), replicating standard timing schemes of a pricing policy against both morning and afternoon peak congestion. We then extend the policy hours to cover a longer period of 12h from 7:00 AM to 7:00 PM (scenario A2). Comparing the two addresses the sensitivity of the timing parameter in the study case by examining how shifting the time window by one hour affects system behavior (e.g., whether changes in inflow and traffic are equal, proportional, or exhibit a different pattern under such shifts). 
Next, we limit the policy to morning hours (8:00 AM to 1:00 PM, scenario A3) and to afternoon hours (1:00 PM to 6:00 PM, scenario A4). Comparing scenarios A3 and A4 allows us to understand whether the policy is more effective in alleviating traffic congestion during the morning peak or the afternoon peak.

Moreover, we incorporate scenarios where the policy targets different users based on their socio-economic characteristics, or given the differences in their vehicle emission levels. The goal of the comparison is to assess the ability of the policy to account for the differences among vehicle or user characteristics (e.g., socio-economic group, vehicle class) and to what extent the relevant indices are impacted by this group-differentiated policy. Focusing on emission reductions, we consider variations in which the price is differentiated by vehicle emission class, with a ticket fee of \euro10 for Euro-class levels 0--3 and a reduced fee of \euro5 for the remaining levels 4--6 (scenario A5). Exploring user exemptions, we differentiate per socio-economic level of the population and exempt users with economic fragility (i.e., yearly income $<$\euro13k), corresponding to 10\% of the population (scenario A6). 

To assess the sensitivity of system outcomes to behavioral assumptions, we compare scenarios in which the same daily fixed-price policy is applied to 100\% of the users while adjusting key behavioral parameters to capture extreme cases. These include a highly-reduced cost acceptability, where the ticket fee is fixed at \euro5 but the acceptable price is reduced to be less than \euro0.5 (scenario B1). Then, we consider the removal of time-adjustment flexibility (scenario B2) and the removal of modal shifts (scenario B3), which prevent considering time- and mode- variability, respectively, as acceptable strategies. These three extreme cases each strongly disincentivize a specific behavioral strategy, thereby allowing us to isolate and analyze the effect of excluding that strategy. This approach is particularly informative because behavioral strategies are difficult to model and their associated parameters are often subject to substantial uncertainty. By examining the extreme case in which a given strategy is absent, we effectively explore a boundary state of the system.

The parameters defining the scenarios A1--A6 are reported in Table \ref{tab:scenariosA}, while parameter configurations of scenarios B1--B3 are reported in Table \ref{tab:scenariosB}. 

\begin{table}[tb]
    \small{
    \centering
    \begin{tabular}{llcccc}
    \toprule
        \textit{ID} & \textit{Description} & $t_s$ & $t_e$ & $F_e$ [\%] & $C_0$ [\euro] \\
    \midrule
         A1 & Basic policy & 8:00 AM & 6:00 PM & 0 & \{5 $\forall l$\}  \\
         A2 & Extended-time policy & 7:00 AM & 7:00 PM & 0 & \{5 $\forall l$\}  \\
         A3 & Morning policy & 8:00 AM & 1:00 PM & 0 & \{5 $\forall l$\}  \\
         A4 & Afternoon policy & 1:00 PM & 6:00 PM & 0 & \{5 $\forall l$\}  \\
         A5 & Emission policy & 8:00 AM & 6:00 PM & 0 & \{10 if $l=0,1,2,3$; 5 otherwise\} \\
         A6 & Exempted policy & 8:00 AM & 6:00 PM & 10 & \{5 $\forall l$\}  \\
    \bottomrule

    \end{tabular}
    \caption{Parameter configuration of the scenarios A1--A6, assessing changes in policy impacts given changes in control parameters. The remaining model parameters are set equal for all the scenarios: among behavioral parameters, $C_{50\%}=4-7$\euro, $\Delta_{a50\%}=\Delta_{p50\%}=1$ hours, $\Delta_{ar50\%}=\Delta_{pr50\%}=1.5$ hours, $\beta_0=-1.24$, $\beta_1=4.5$, $\beta_2=-1.45$, $\beta_3 = -0.30$, $\beta_4=-0.034$; for the dwell time distribution, $\bar{\tau}=20$ minutes; for the emission conversion factor, $km\nu=2.5$.}
    \label{tab:scenariosA}
    }
\end{table}

\begin{table}[tb]
    \small{
    \centering
    \begin{tabular}{llcccc}

    \toprule
        \textit{ID} & \textit{Description} & $C_{50\%}$ [\euro] & $\Delta_{a50\%} [h]$ & $\Delta_{p50\%}[h]$ & Shift to PTS \\
    \midrule
         B1 & No economic-flexibility & 0.1--0.5 & 1.5 &  1.5 & True  \\
         B2 & No time-flexibility & 4--7 & 0 & 0 & True  \\
         B3 & No mode-flexibility & 4--7 & 1.5 & 1.5 & False  \\
    \bottomrule
    \end{tabular}
    \caption{Parameter configuration of the scenarios B1--B3, assessing changes in policy impacts given changes in behavioral parameters. The remaining model parameters are set equal for all the scenarios: among control parameters, $t_s=$ 8:00 AM, $t_e=$ 6:00 PM, $C_0=5$\euro, $F_e=0\%$; among other behavioral parameters, $\Delta_{ar50\%}=1.5$ hours, $\Delta_{pr50\%}=1.5$ hours, $\beta_0=-1.24$, $\beta_1=4.5$, $\beta_2=-1.45$, $\beta_3 = -0.30$, $\beta_4=-0.034$; for the dwell time distribution, $\bar{\tau}=20$ minutes; for the emission conversion factor, $km\nu=2.5$.}
    \label{tab:scenariosB}
    }
\end{table}

\subsubsection{Evaluation setup}\label{subsub:evaluation}

The comparison between scenarios, which is reported in Subsection~\ref{subsub:findings}, covers both qualitative and quantitative analysis. First, we concentrate on a set of global metrics that summarize overall system behavior. They include the total number of vehicles entering the area per day (i.e., daily inflow), the maximum number of vehicles circulating in the area during the day (i.e., max daily traffic), and the maximum number of vehicles circulating in the area during the policy activation hours (i.e., max traffic in policy hours). Note that these quantities differ in the measurement units: while inflows measure the cumulative number of entering vehicles during the whole day, the maximum traffic accounts for the maximum number of vehicles circulating at the same time in the interest period. Moreover, the analysis considers the total grams of NOx emitted by the vehicles circulating in the area during the day (i.e., daily emissions). Other than these key mobility indicators, other metrics are reported to better describe the what-if scenarios, such as the absolute number of vehicles changing their habits with a time-shift, with a modal-shift, or in another way (i.e., lost). Finally, the total earnings collected with the access tickets are summarized in the global indicator of daily revenue. 

In addition to aggregated values, we present the temporal evolution of these metrics in graphical representations. The visualizations enable us to observe how system dynamics evolve and identify patterns that may not be evident from aggregate values alone. In the following sections, we present time series plots representing inflow, traffic, and emissions within the area of interest throughout the day. The x-axis represents time, while the y-axis displays the indicator of interest. Different colors denote the various scenarios under comparison, enabling a clear visual distinction between policy implementations.

\subsubsection{Findings}\label{subsub:findings}

For the proposed scenarios A1--A6, Table \ref{tab:global} collects the measured values of the global indicators. As expected, the results demonstrate that when access policies are activated, there is a significant reduction in inbound flows, traffic, and emissions. It is particularly interesting to quantify how the intensity of these reductions varies depending on the applied policy.
Scenarios A1--A4 highlight how the policy duration significantly affects traffic and emissions. Scenario A2 results in a larger reduction in daily inflow with respect to A1 (-56.5k vs -44.6k) and in traffic during policy hours (-5.0k vs -1.5k), with a more pronounced decrease in emissions (-209.8kg NOx vs -166.7kg NOx), but also a substantial reduction in maximum traffic during the day (-1.5k) in contrast to the increase in the A1 scenario (+1.4k). Scenarios A3 and A4 illustrate the effect of timing: the morning policy in A3 produces smaller impacts on inflow (-17.0k) and traffic (-0.6k policy-hours), whereas the afternoon policy in A4 modifies daily maximum traffic moderately (+0.9k) and maximum traffic in policy hours significantly (-4.4k), with emissions (-10.1\% vs -11.9\%) and revenue (112k in both scenarios) effects more balanced.

Regarding policies tailored for Euro-class vehicles or including exemptions, daily inflow reductions range from -51.5k in A5 to -38.4k in A6, with A1 in between (-44.6k), while variations in traffic during policy hours are minor (-1.5k for A5, -1.4k for A6, -1.5k for A1). Emissions reductions follow the same pattern, from -192.8k NOx in A5 to -143.7k NOx in A6. Time- and mode-shifted vehicles, as well as lost vehicles, also show consistent trends across the three scenarios. Overall, the results indicate that despite slight numerical differences, the three scenarios produce broadly comparable impacts on system performance and revenue.

Focusing on the changes in the behavioral assumptions, the global indicators relative to scenarios B1--B3 are reported in Table \ref{tab:global2}. All scenarios see a reduction in daily inflows and NOx emissions, but the impact on peak traffic and user behavior differs significantly. Scenario B1 shows the largest reductions in inflows (-43\%) and emissions (-44\%), but peak traffic slightly increases due to time-shifting, suggesting that overall demand reduction is not sufficient to prevent congestion during critical hours. The very low revenues ($>$\euro700/day) indicate minimal financial uptake, despite substantial adjustments by a subset of users. Scenario B2 achieves a moderate decrease in inflows (-32\%) and emissions (-32\%), with a clear reduction in peak traffic (-19\% in policy hours). This indicates that rigidity and mode-shift alone can effectively mitigate congestion while maintaining significant revenue (\euro252.5k/day). Finally, scenario B3 shows limited environmental gains (\text{-15}\% in emissions) and increased peak traffic, reflecting the concentration of demand in fewer hours. Revenues are highest (\euro271.8k/day), but congestion and lost trips remain substantial.

\begin{table}[tb]
    \centering
    \footnotesize{
    \begin{tabular}{lcccccc}
    \toprule
        & \multicolumn{6}{c}{\textit{Scenario}} \\
        \textit{Indicator}  & A1 & A2 & A3 & A4 & A5 & A6 \\
    \midrule
        Daily inflow [veh]  
            & -44.6k & -56.5k & -17.0k & -19.6k & -51.5k & -38.4k \\
            & (-26.5\%) & (-33.6\%) & (-10.1\%) & (-11.6\%) & (-30.7\%) & (-22.9\%) \\
        Max traffic in the day [veh]  
            & 1.4k & -1.5k & 1.2k & 0.9k & 1.7k & 1.1k \\
            & (9.9\%) & (-10.7\%) & (8.3\%) & (6.5\%) & (11.9\%) & (7.6\%) \\
        Max traffic in the policy hours [veh]  
            & -1.5k & -5.0k & -0.6k & -4.4k & -1.5k & -1.4k \\
            & (-10.2\%) & (-35.3\%) & (-4.7\%) & (-31.1\%) & (-10.2\%) & (-9.8\%) \\
        Daily emissions [g NOx] 
            & -166.7k & -209.8k & -62.7k & -73.8k & -192.8k & -143.7k \\
            & (-26.8\%) & (-33.8\%) & (-10.1\%) & (-11.9\%) & (-31.0\%) & (-23.1\%) \\
        Time-shifted [veh] 
            & 46.2k & 43.6k & 37.4k & 42.2k & 51.2k & 40.4k \\
        Mode-shifted [veh] 
            & 86.5k & 106.9k & 36.5k & 41.3k & 95.0k & 75.7k \\
        Lost [veh] 
            & 59.0k & 75.4k & 18.5k & 22.2k & 73.3k & 49.6k \\
        Daily revenue [\euro] 
            & 230k & 293k & 112k & 112k & 255k & 217k \\
            
     \bottomrule
    \end{tabular}
    \caption{Comparison of global indicators in the scenarios A1--A6 with respect to the baseline situation.}
    \label{tab:global}
    }
\end{table}

\begin{table}[tb]
    \centering
    \small{
    \begin{tabular}{lccc}
    \toprule
        & \multicolumn{3}{c}{\textit{Scenario}} \\
        \textit{Indicator}  & B1 & B2 & B3 \\
    \midrule
        Daily inflow variation [veh]  & -72.9k (-43\%) & -53.7k (-32\%) & -24.5k (-15\%) \\
        Max traffic variation in the day [veh]  & 5.2k (0.36\%) & -1.1k (-0.08\%) & 3.9k (0.28\%) \\
        Max traffic variation in policy hours [veh] & 291 (2.0\%) & -2.7k (-19\%) & 529 (3.7\%) \\
        Daily emissions variation [g NOx] & -273.3k (-44\%) & -199.9k (-32\%) & -91.9k (-15\%) \\
        Time-shifted [veh] & 103.9k & 2.8k & 84.7k \\
        Mode-shifted [veh] & 115.3k & 95.9k & 0 \\
        Lost [veh] & 121.8k & 81.1k & 77.2k \\ 
        Daily revenue [\euro] & $>$500 & 252.5k & 271.8k \\
     \bottomrule
    \end{tabular}
    \caption{Comparison of global indicators in the scenarios B1--B3 with respect to the baseline situation.}
    \label{tab:global2}
    }
\end{table}

\begin{figure}[tb]
    \centering
    \includegraphics[width=0.32\textwidth]{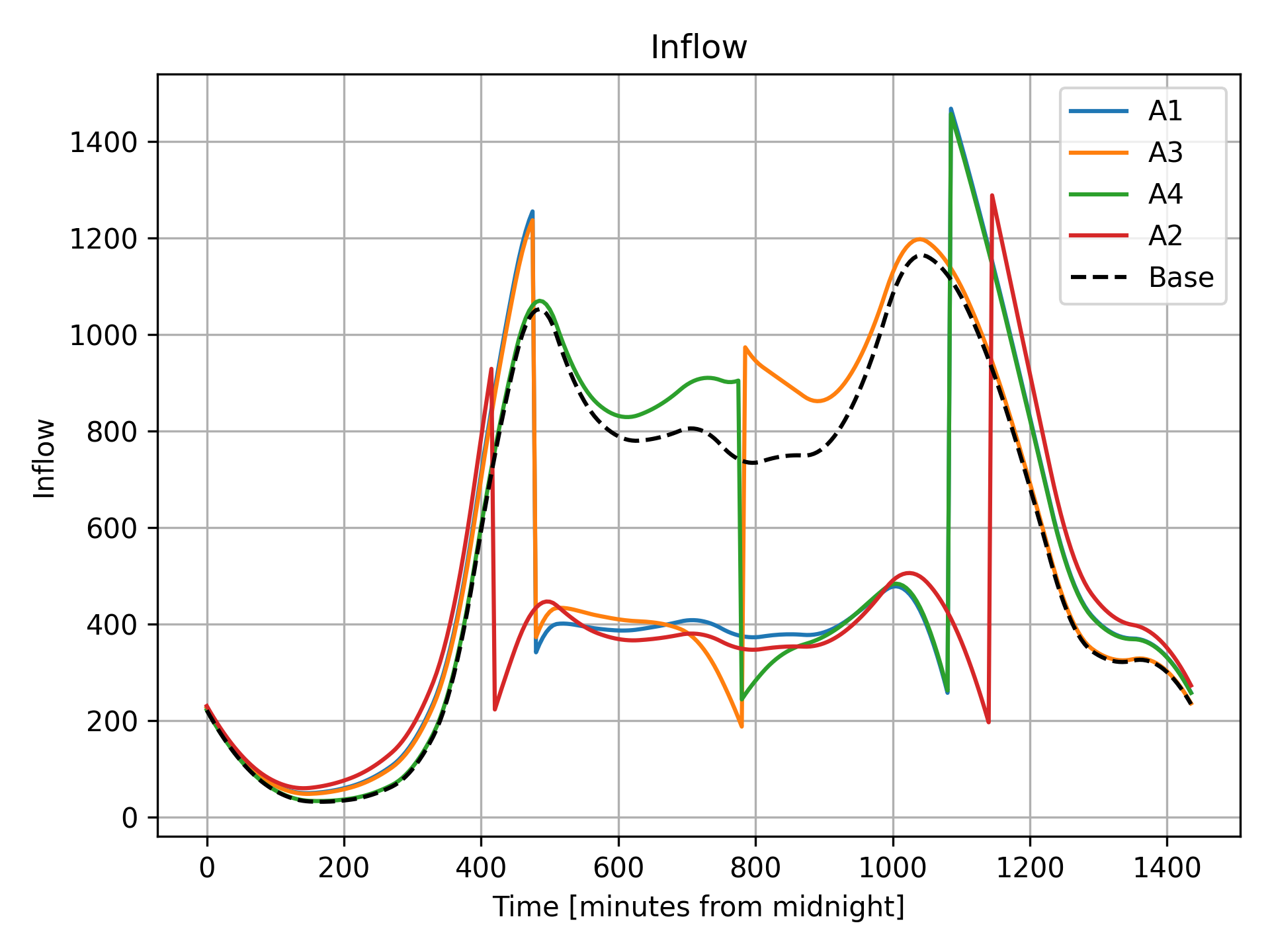}\hfill
    \includegraphics[width=0.32\textwidth]{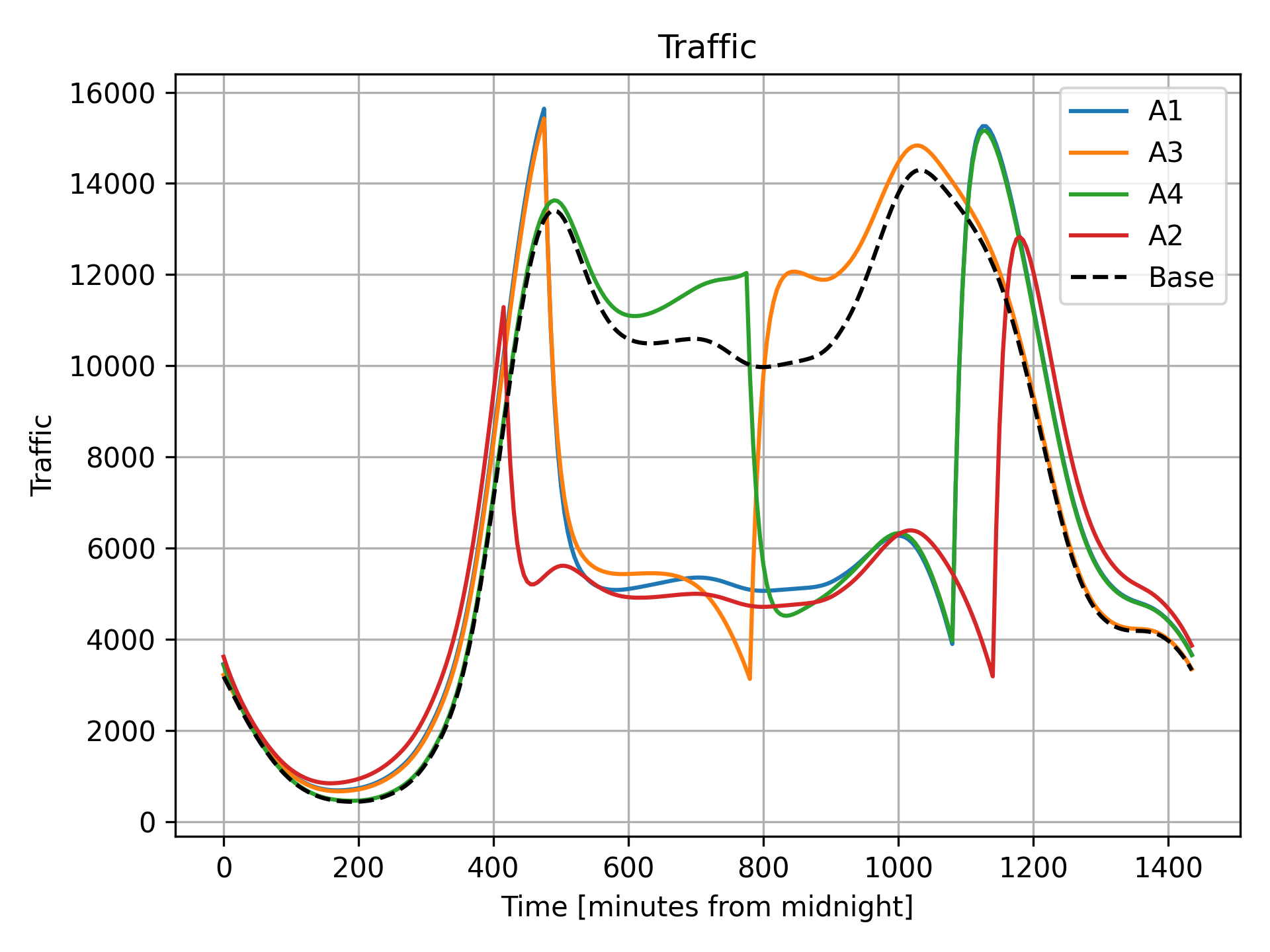}\hfill
    \includegraphics[width=0.32\textwidth]{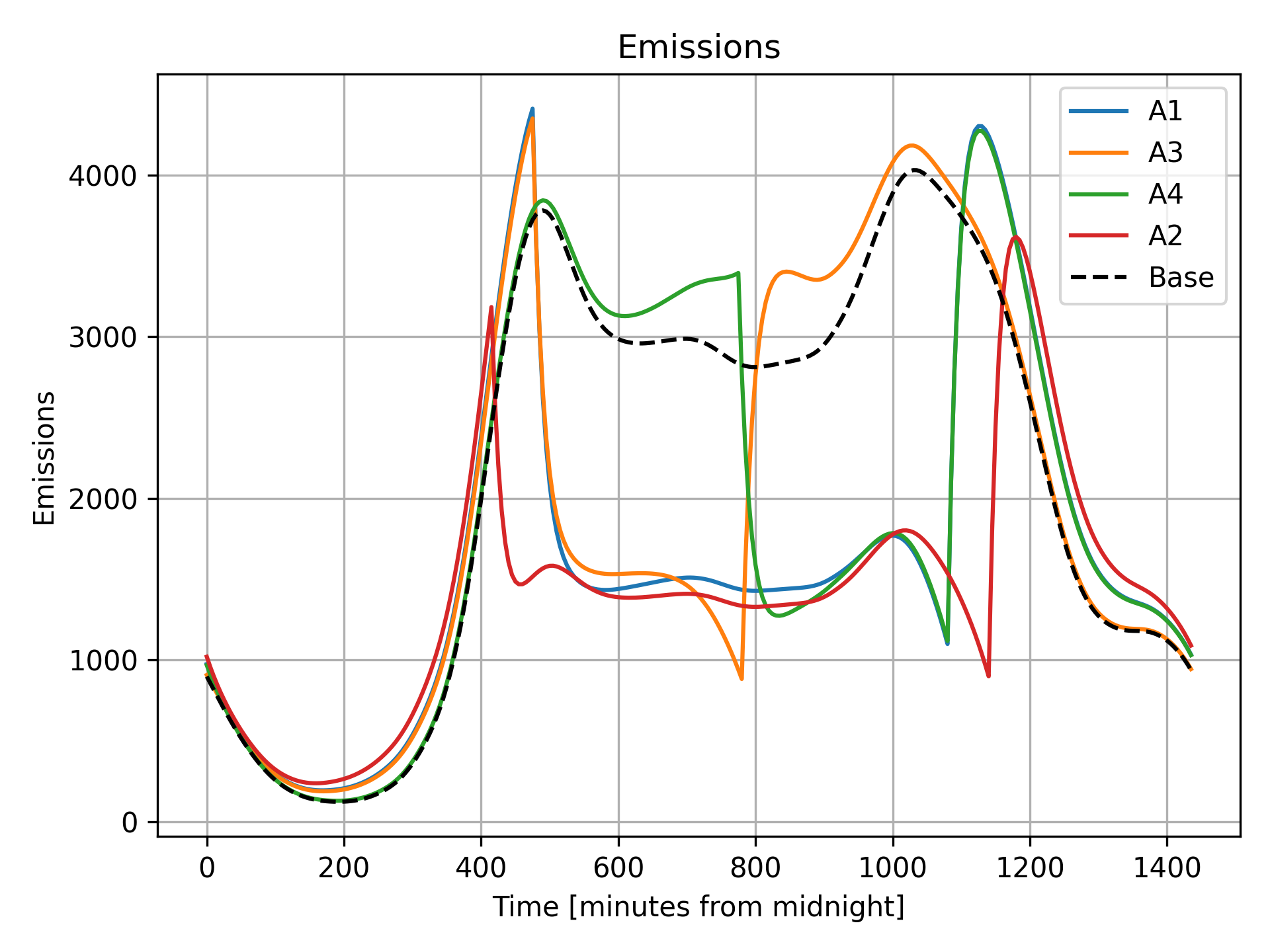}
    \caption{Inflow [veh], traffic [veh], and emissions [g NOx] varying over time for the scenarios A1--A4.}
    \label{fig:kpi_plots_A1}
\end{figure}

\begin{figure}[tb]
    \centering
    \includegraphics[width=0.32\textwidth]{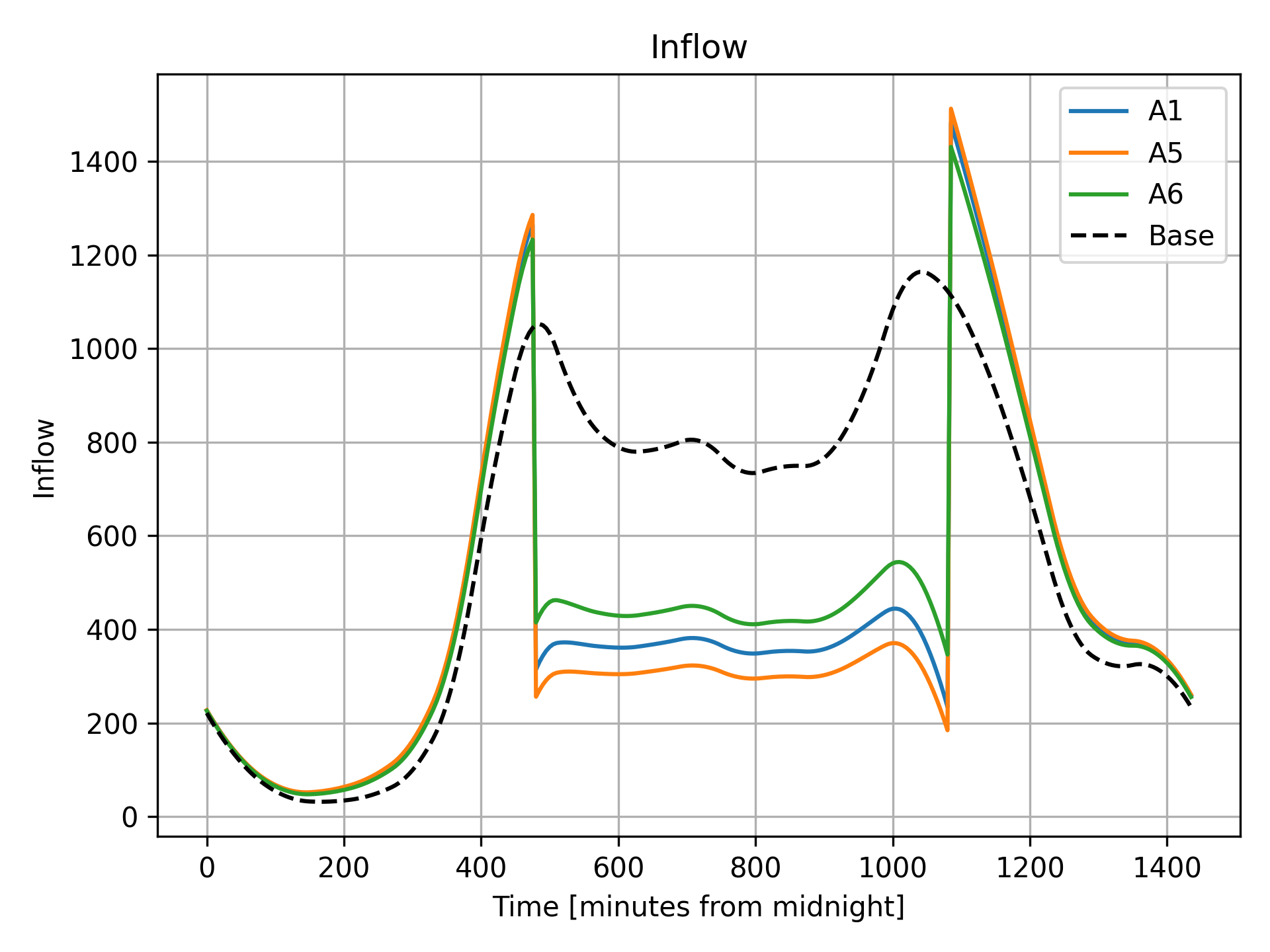}\hfill
    \includegraphics[width=0.32\textwidth]{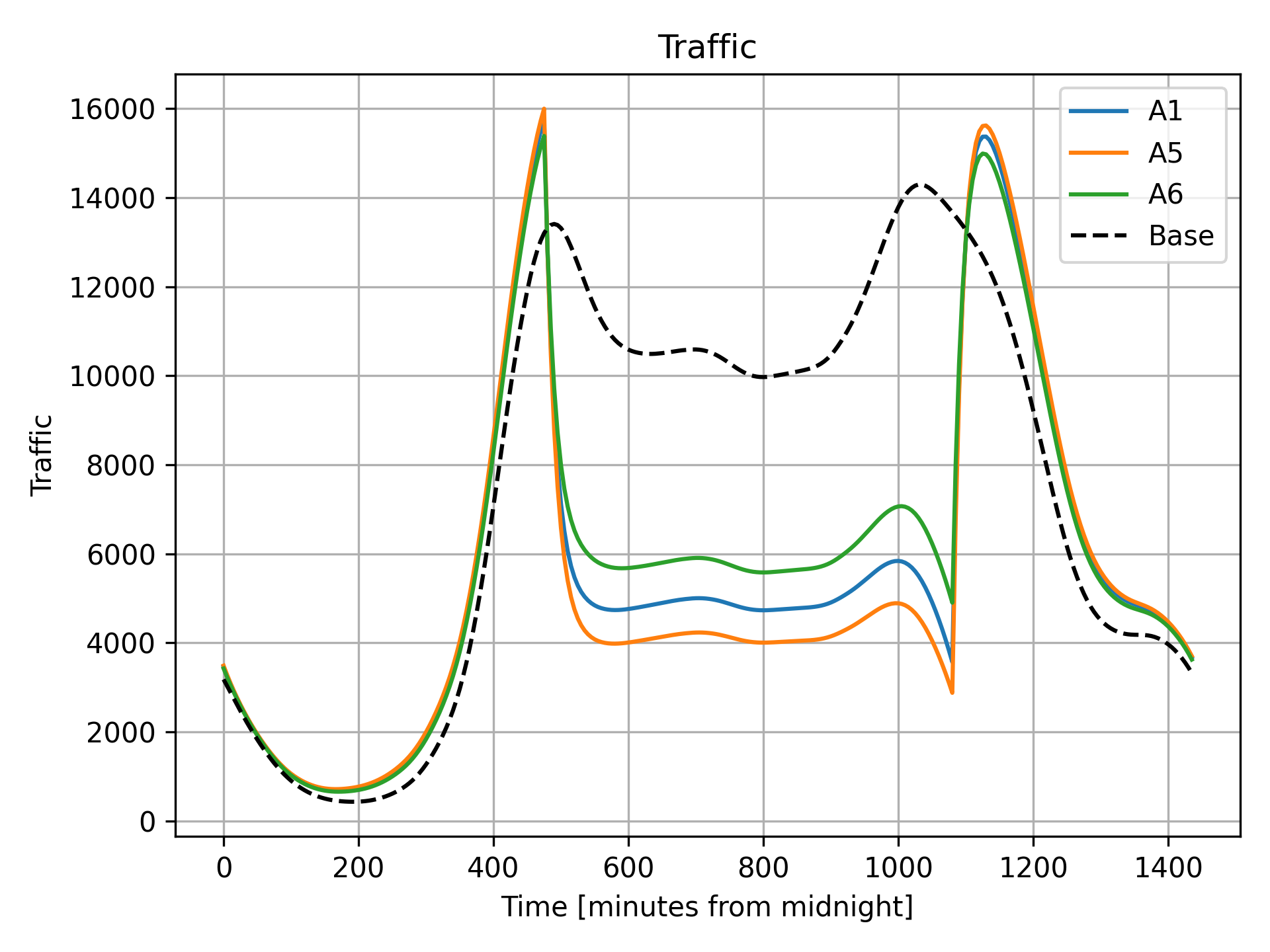}\hfill
    \includegraphics[width=0.32\textwidth]{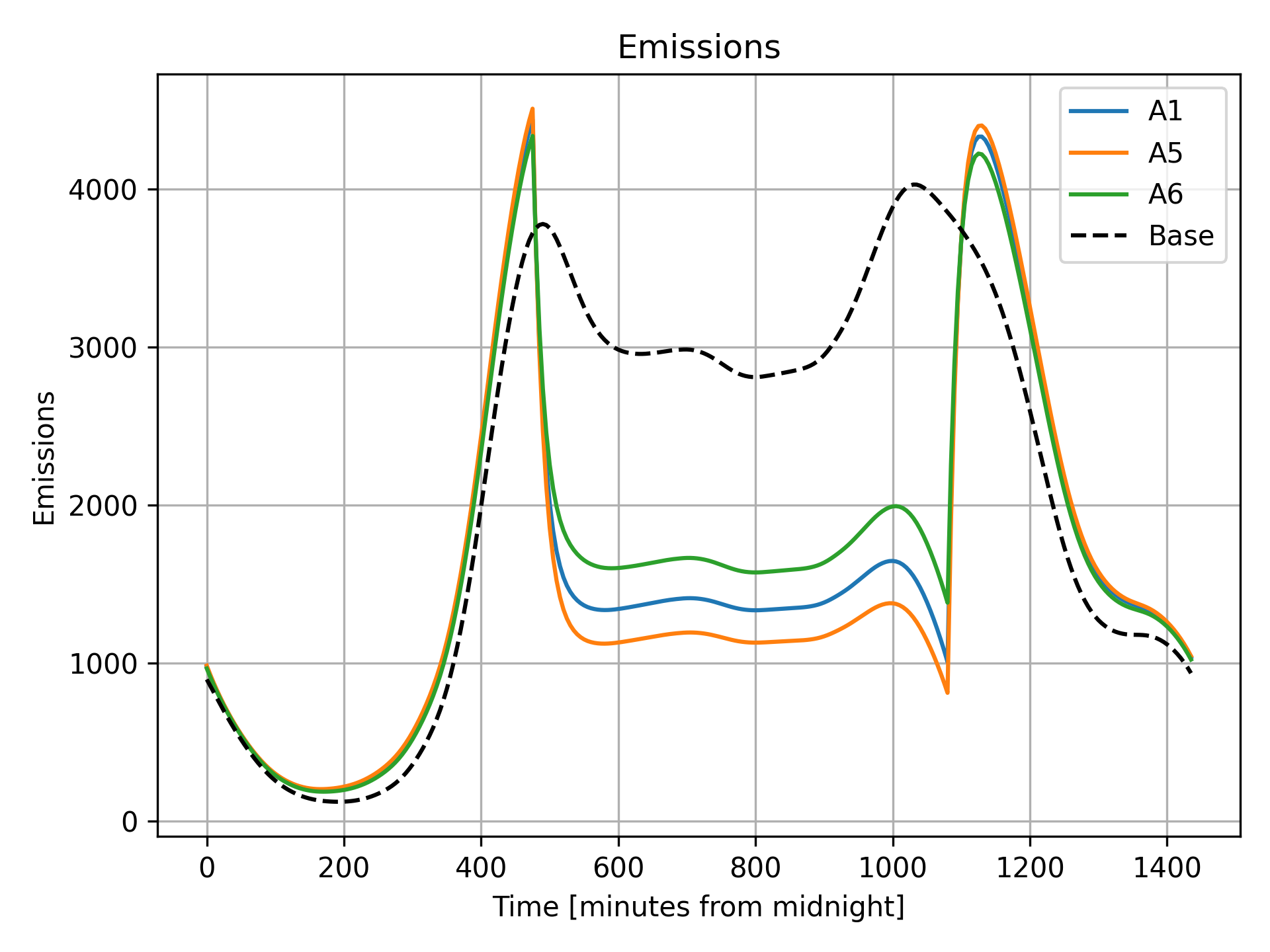}
    \caption{Inflow [veh], traffic [veh], and emissions [g NOx] varying over time for the scenarios A1, A5--A6.}
    \label{fig:kpi_plots_A2}
\end{figure}

\begin{figure}[tb]
    \centering
    \includegraphics[width=0.32\textwidth]{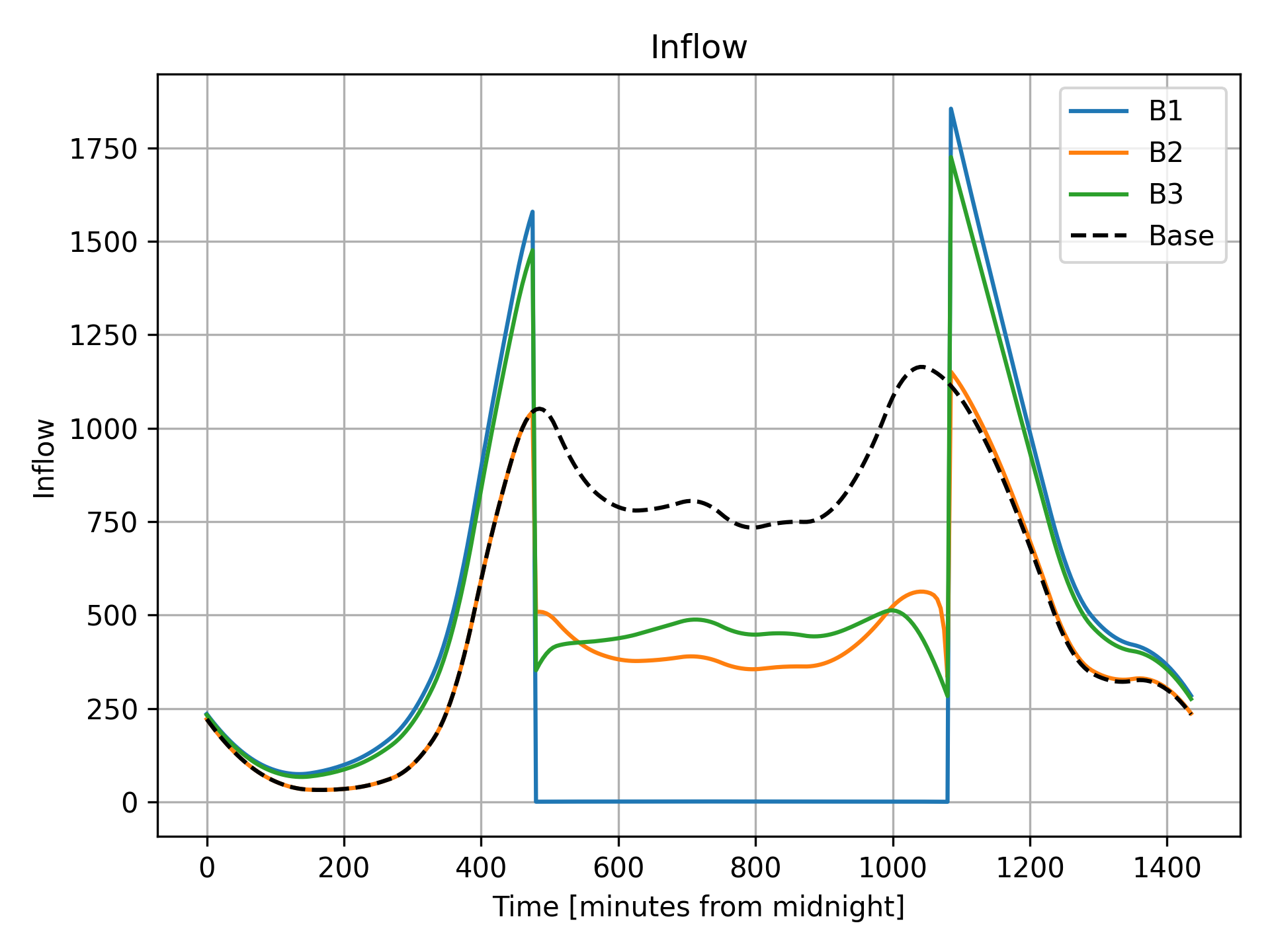}\hfill
    \includegraphics[width=0.32\textwidth]{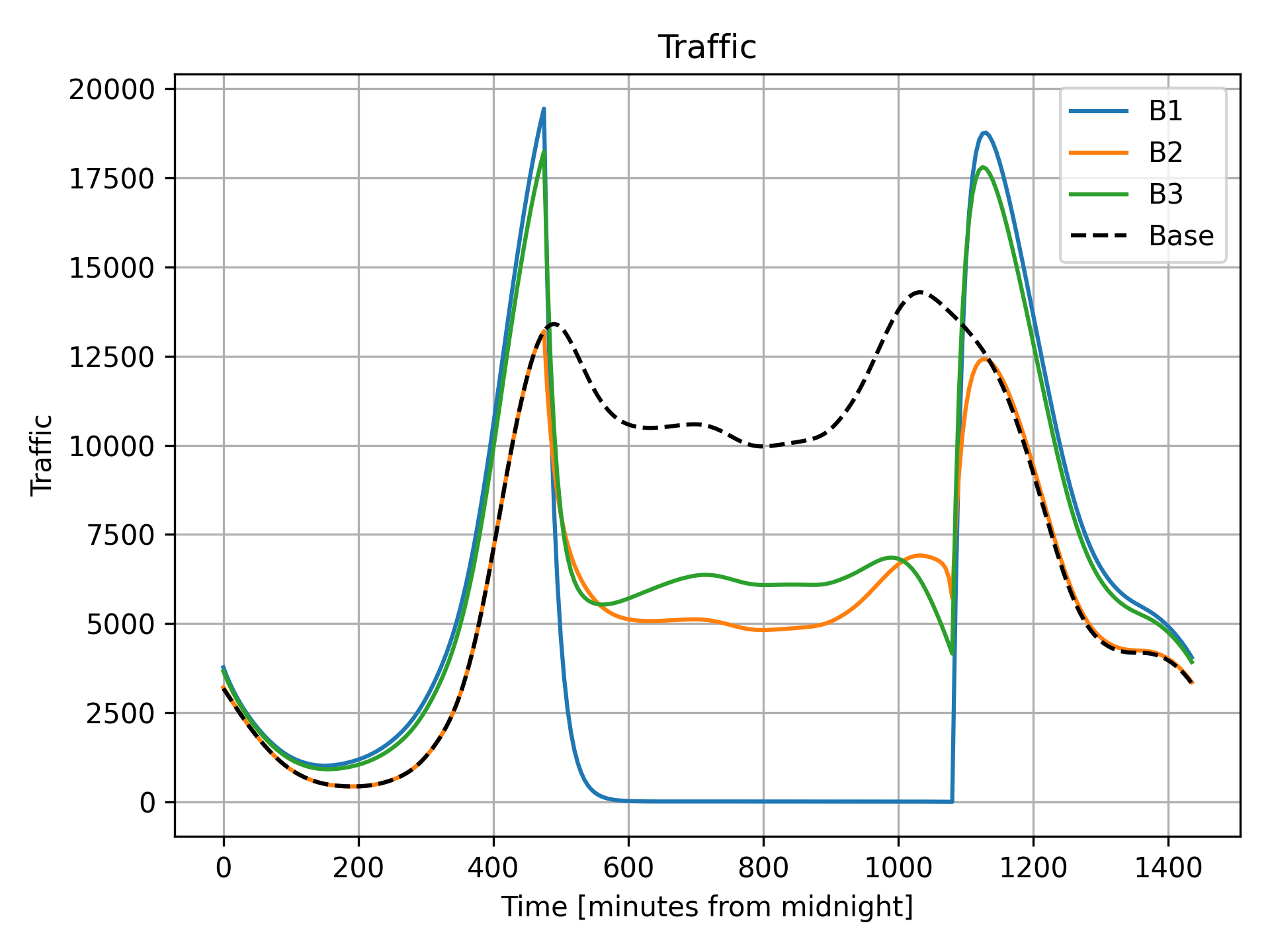}\hfill
    \includegraphics[width=0.32\textwidth]{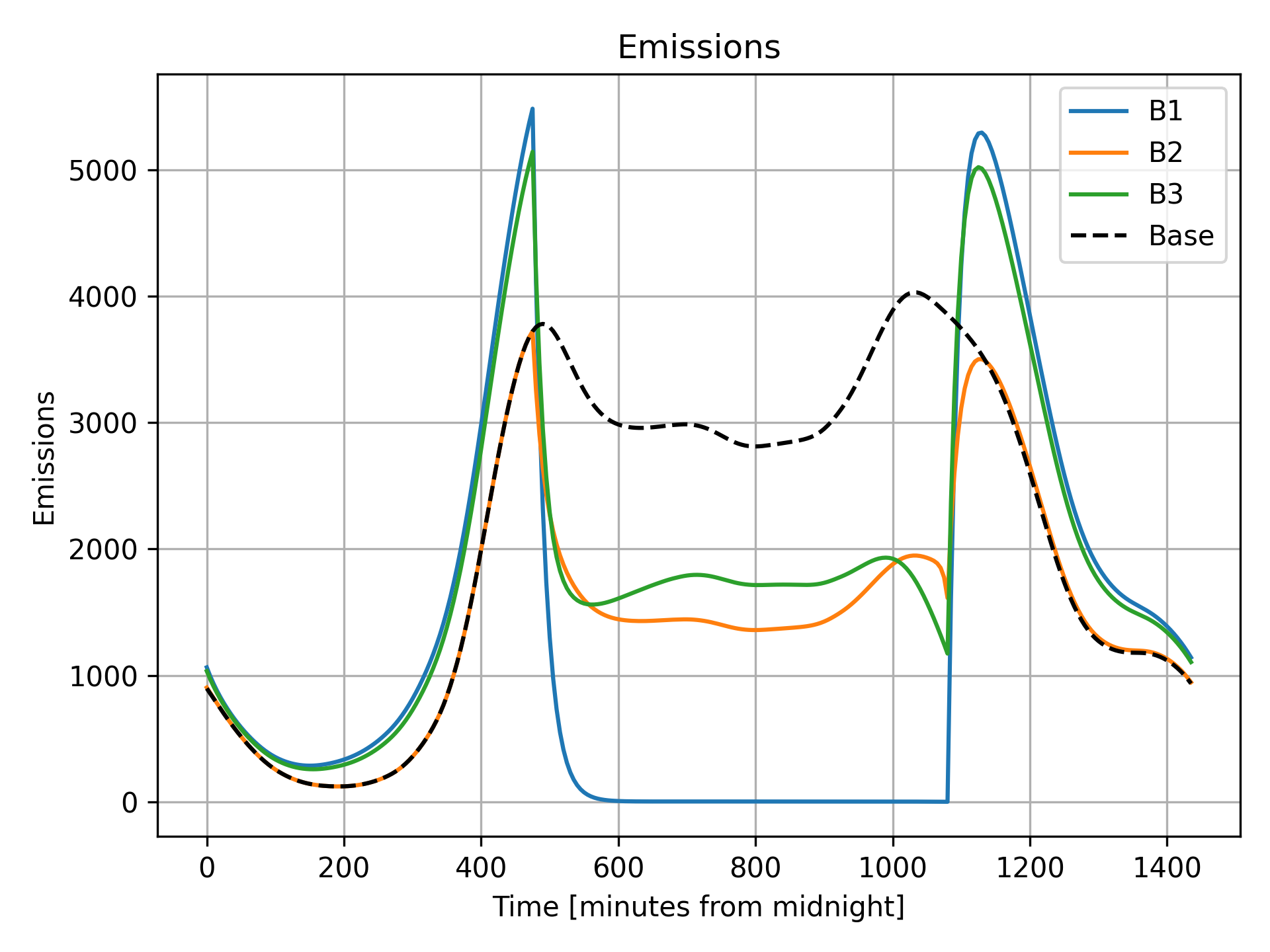}
    \caption{Inflow [veh], traffic [veh], and emissions [g NOx] varying over time for the scenarios B1--B3.}
    \label{fig:kpi_plots_B}
\end{figure}

The qualitative analysis throughout visualizations is shown in Figures \ref{fig:kpi_plots_A1}, \ref{fig:kpi_plots_A2}, and \ref{fig:kpi_plots_B}.
The higher daily maximum traffic in A1 compared to A2 is driven by the combined effect of vehicles anticipating or postponing trips, adding to an already high congestion level, whereas traffic remains more moderate throughout A4. Similarly, A1, A3, and A4 all experience peaks due to anticipatory and delayed trips, generating local surges in flow. Notably, morning traffic is lower overall, which explains why A3 shows smaller total reductions in daily inflow compared to A4. These patterns highlight how both the timing and duration of the policy shape daily traffic dynamics, congestion peaks, and the environmental effectiveness, beyond their impact on total demand.

Figure \ref{fig:kpi_plots_A2} highlights the differences between scenarios A1, A5, and A6, which are present but largely homogeneous over time rather than exhibiting strong temporal patterns. This indicates that, in this case, global indicators are sufficient to capture and analyze the impacts of the scenarios, without requiring detailed time-resolved examination.

Figure \ref{fig:kpi_plots_B} illustrates the distribution of the quantities over time for scenarios B1--B3. The results reveal how behavioral constraints fundamentally reshape traffic patterns under mobility pricing. In scenario B1, traffic, inflow, and emissions become highly volatile with extreme peaks as travelers concentrate trips to avoid fees, followed by near-zero periods in policy activation hours. Scenario B2, on the other hand, produces the most stable outcomes, with sustained moderate flows and emissions, and no high peaks near policy hours. The inability to shift from car to bus (scenario B3) generates an intermediate pattern with significant peaks close to policy activation hours, but maintained baseline traffic within policy hours. The examination of these scenarios highlights the strong dependence of system-level outcomes on behavioral flexibility assumptions, supporting the importance of comprehensive sensitivity analysis. In particular, small changes in travelers’ ability to reschedule trips or switch modes can lead to qualitatively different temporal traffic profiles and emission dynamics. These findings underscore that policy evaluations based on a single behavioral specification may be misleading and that robust mobility policy assessment requires explicit consideration of heterogeneous traveler responses.

\begin{figure}[tb]
  \centering
    \includegraphics[width=0.32\textwidth]{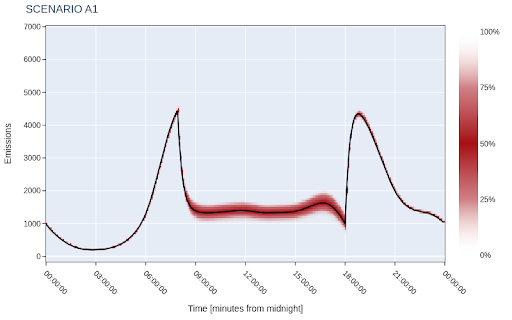}\hfill
    \includegraphics[width=0.32\textwidth]{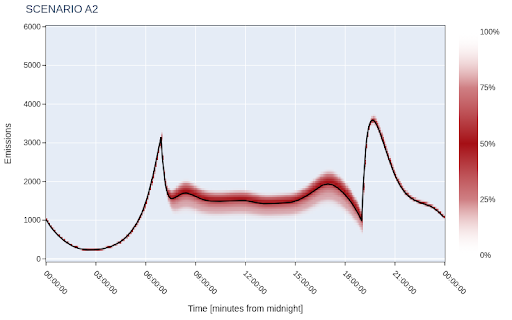}\hfill
    \includegraphics[width=0.32\textwidth]{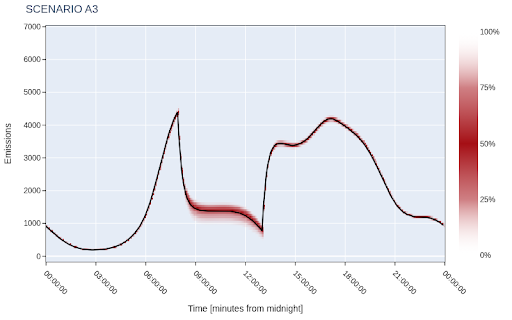}\\
    \includegraphics[width=0.32\textwidth]{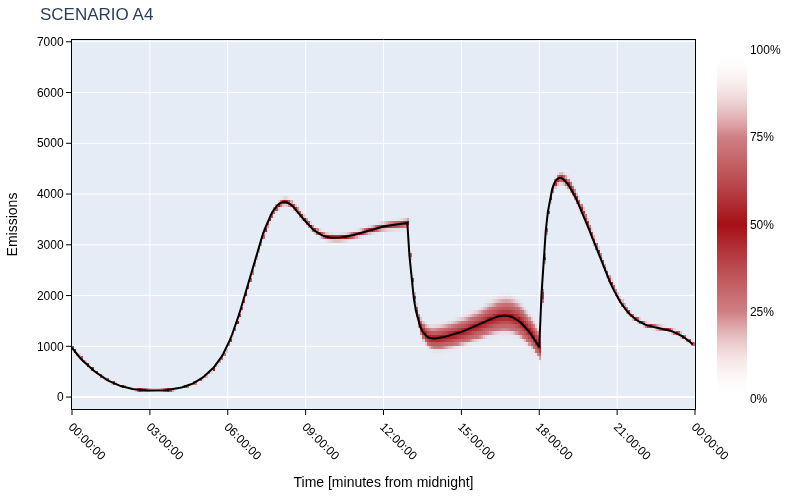}\hfill
    \includegraphics[width=0.32\textwidth]{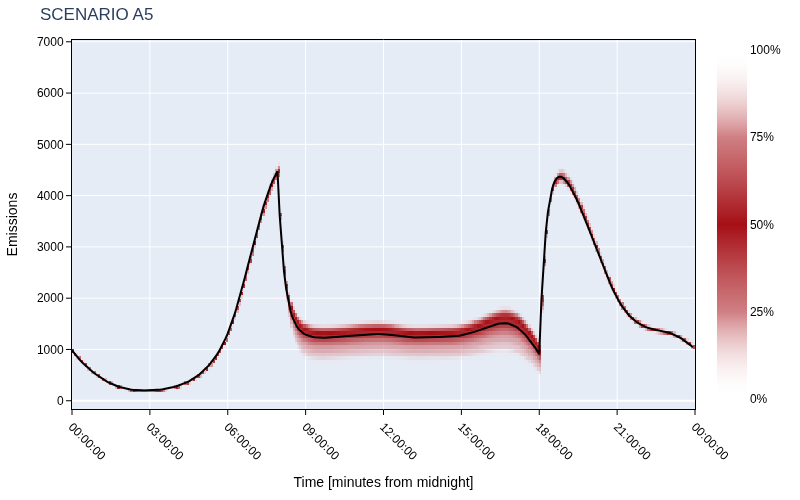}\hfill
    \includegraphics[width=0.32\textwidth]{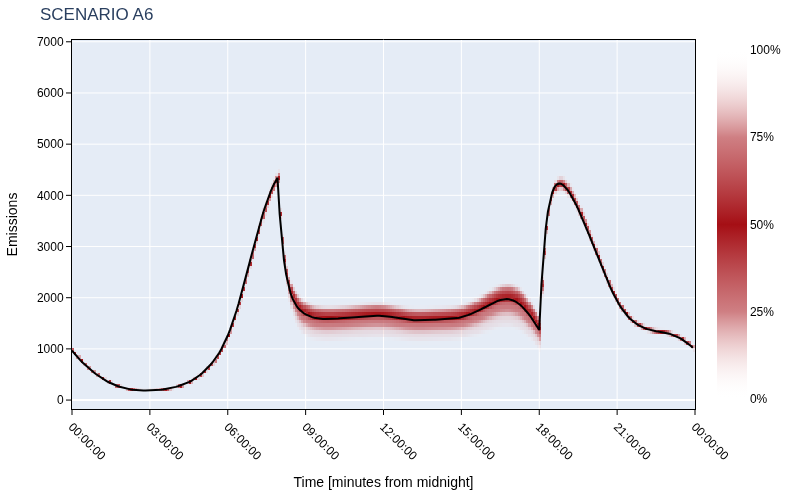}\\
    \includegraphics[width=0.32\textwidth]{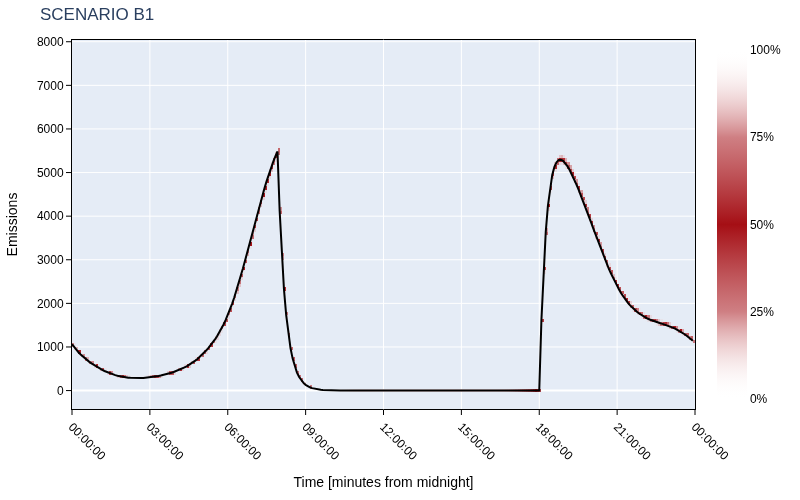}\hfill
    \includegraphics[width=0.32\textwidth]{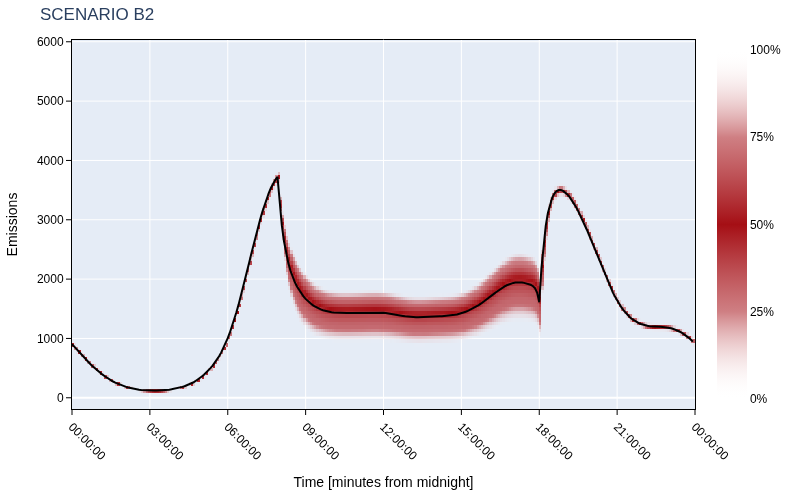}\hfill
    \includegraphics[width=0.32\textwidth]{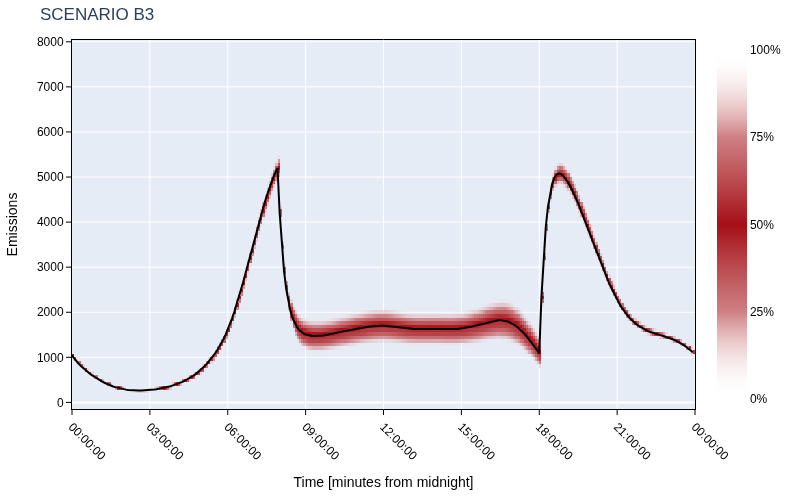}
  \caption{Uncertainty bands of emission estimates [g NOx] over time for scenarios A1–A6 and B1–B3 (ordered left-to-right, top-to-bottom). Color density represents estimation confidence using a divergent scale: the central value (50), shown in red, indicates maximum confidence (the most probable value), while the color fades to white toward the boundaries (0 and 100) to signify decreasing confidence.}
  \label{fig:uncertainty_plot}
\end{figure}

We remark that, for the sake of simplicity in the comparison among different what-if scenarios, the analysis of the expressiveness done so far does not cover the uncertainty in the predictions. However, all the analyses can be extended to this aspect as well as prediction uncertainty is a built-in feature of the model. Figure \ref{fig:uncertainty_plot} shows the emissions for scenarios A1–A6 and B1–B3 already discussed, but enriched with uncertainty bands. Unsurprisingly, the uncertainty is higher during the hours when the policy is active. During these hours, the uncertainty depends on the predictability of the changes in the behaviors. For instance, in the case of B1, the uncertainty is low since the scenario imposes extreme sensitivity to the policy costs.

\section{Conclusions} \label{sec:conclusions}

In this study, we present a model-driven, simulation-based approach for the ex-ante evaluation of urban traffic regulation policies, capable of capturing both direct impacts, such as traffic conditions, and indirect impacts, including environmental effects related to mobility. The application of the framework to the real-world case study of Bologna demonstrates that the integration of a physical layer, representing the transportation system and traffic flows, with a social layer, capturing individual behaviors and adaptation to policy interventions, enables systematic exploration of what-if scenarios and assessment of the potential consequences of alternative regulatory designs before their actual implementation.
The proposed approach addresses the complexity inherent in socio-technical urban mobility systems, accounting for nonlinear interactions, uncertainty, and dynamic adaptation of user behavior, while supporting transparent and reproducible generation of counterfactual scenarios. In this way, it enables a rigorous assessment of traffic policies before their implementation, offering decision-makers actionable insights into potential impacts.

The evaluation results highlight the analytical strength of the proposed framework. In particular, the framework exhibits the capacity to discriminate between policy configurations that are structurally similar, showing when differences in policy design produce meaningfully different effects on the system and when they lead to similar overall outcomes. Moreover, the analysis shows that global indicators are not always sufficient to fully characterize policy effects: while in some cases aggregate measures provide an adequate assessment, in others a temporal examination is necessary to correctly interpret congestion and flow patterns. This ability to combine aggregate and temporal perspectives strengthens the robustness of the policy evaluation process. Finally, although uncertainty is not the primary focus of this study, the framework naturally accommodates it, enabling future extensions toward fully uncertainty-aware assessments.
In addition, experimenting with extreme scenarios, in particular for what concerns behavioral parameters, appears to be a powerful approach to assess the sensitivity to these parameters: this sensitivity analysis is particularly relevant for the evaluation of the potential impacts of mobility policies, given the difficulty to provide precise and accurate models for behavioral parameters.

Several directions could enhance the proposed framework. 
From a spatial perspective, the current model works at a single-zone level and provides aggregated insights for one urban area. A natural extension would involve a multi-scale approach that divides the selected area into finer spatial units. This would enable the evaluation of the same key indicators at a sub-zone level. Such granularity would reveal localized hotspots and spatial patterns that may be hidden by aggregation, supporting more targeted interventions and allowing policymakers to design zone-specific measures.

From a temporal perspective, the current model assumes a short-term perspective with independence in individual choices. Extending the temporal horizon would allow policymakers to observe how the urban mobility system evolves over time in response to policy interventions. This would provide insights into the long-term effects of mobility plans and support more informed strategic planning.

Concerning the social dimension, the proposed framework currently focuses on efficiency metrics such as inflows, traffic congestion, and emissions. However, urban mobility policies have different impacts across socio-economic groups. Future work should incorporate indicators of social vulnerability and equity to assess how policy interventions affect disadvantaged communities. By combining efficiency with equity considerations, the model would provide a more complete evaluation framework aligned with principles of environmental fairness and inclusive urban planning.

\paragraph{\textit{Aknowledgements:} }

The authors acknowledge the Municipality of Bologna for providing both the data and practical expertise. We also thank Thomas Louf for his contribution in estimating the parameters used in the emissions calculations.

\paragraph{\textit{Fundings:} }

This work was supported by project ``Gemello Digitale: Governo e Valorizzazione del Patrimonio Dati'' (project BO1.1.2.1.a) in the scope of the National Metro Plus Program and Southern Medium Cities 2021-2027 and the ``ICSC National Research Centre for High Performance Computing, Big Data and Quantum Computing'' (CN00000013), under the NRRP MUR program funded by the NextGenerationEU.

\bibliographystyle{plainnat}
\footnotesize{
\bibliography{references}
}

\end{document}

%% file: schemas/schema1.tex
\begin{figure}
    \centering
    \resizebox*{10cm}{!}{
\begin{tikzpicture}[
    node distance=2cm,
    font=\small,
    block1/.style={rectangle, draw, fill=lightgreenn, minimum width=3cm, minimum height=1cm},
    block2/.style={rectangle, draw, fill=white, minimum width=3cm, minimum height=1cm},
    block3/.style={diamond, draw, fill=darkbluee, text=white, aspect=2, minimum width=2.2cm, text width=2cm, align=center, aspect=2}
]

    \node[block1] (asis) {As-is Scenario};
    \node[block3, right=1cm of asis] (model) {Model};
    \node[block2, above=1cm of model.west] (ctrl) {Control and Behavioral Parameters};
    \node[block1, right=1cm of model] (whatif) {What-if Scenario};
    \node[block3, right=1cm of whatif] (impact) {Impact Evaluation};

    \draw[-Stealth, thick] (asis) -- (model);
    \draw[-Stealth, thick] (ctrl.south) -| (model.west);
    \draw[-Stealth, thick] (model) -- (whatif);
    \draw[-Stealth, thick] (whatif) -- (impact);
    \draw[-Stealth, thick] (asis.south) -- ++(0,-1) -| (impact.west);

\end{tikzpicture}
}
    \caption{Schematic representation of the architectural framework for what-if scenario analysis.}
    \label{fig:intro:scheme}
\end{figure}

%% file: schemas/schema2.tex
\begin{figure}
    \centering
    \resizebox*{10cm}{!}{
\begin{tikzpicture}[
    node distance=1.5cm,
    font=\small,
    block1/.style={rectangle, draw, fill=lightgreenn, minimum width=2cm, minimum height=1cm},
    block2/.style={rectangle, draw, fill=white, fill opacity=0.6, text opacity=1, minimum width=3cm, minimum height=1cm},
    block3/.style={diamond, draw, fill=darkbluee, minimum width=1cm, aspect=1},
    block4/.style={rectangle, fill=white, fill opacity=0.6, text opacity=1, minimum width=3cm, minimum height=1cm},
]
    \node[block1, rounded corners=10pt] (traffic1) {Traffic};
    \node[block1, rounded corners=10pt, above=0.5cm of traffic1] (inflow1) {Inflow};
    \node[block1, rounded corners=10pt, below=0.5cm of traffic1] (emissions1) {Emissions};
    \node[block3, right=1.5cm of traffic1] (m1) {};
    \node[block1, rounded corners=10pt, right=1.5cm of m1] (inflow2) {Inflow};
        \node[block3, right=0.5cm of inflow2] (m2) {};
    \node[block1, rounded corners=10pt, right=0.5cm of m2] (traffic2) {Traffic};
    \node[block3, right=0.5cm of traffic2] (m3) {};
    \node[block1, rounded corners=10pt, right=0.5cm of m3] (emissions2) {Emissions};
    \node[block2, below=2.5cm of inflow2] (control) {Control Parameters};
    \node[block2, right=2cm of control] (behavioral) {Behavioral Parameters};
    
    \node[block4, above=0.5cm of inflow2] (fake1) {};
    \node[block4, below=0.5cm of emissions2] (fake2) {};
    \node[block4, above=0.25cm of inflow1] (title1) {As-is Scenario};
    \node[block4, right=6.5cm of title1] (title2) {What-if Scenario};

    \node[block2, above=0.7cm of inflow2.west, rounded corners=10pt, minimum height=0.5cm, minimum width=0.5cm, text width=1cm, font=\tiny, align=center] (oii1) {other indirect impacts};
    \node[block2, above=0.7cm of traffic2.west, rounded corners=10pt, minimum height=0.5cm, minimum width=0.5cm, text width=1cm, font=\tiny, align=center] (oii2) {other indirect impacts};
    \node[block2, above=0.7cm of emissions2.west, rounded corners=10pt, minimum height=0.5cm, minimum width=0.5cm, text width=1cm, font=\tiny, align=center] (oii3) {other indirect impacts};
    
        \node[
            rectangle, 
            draw=lightgreenn, 
            inner sep=15pt, 
            fit=(inflow1) (traffic1) (emissions1) 
        ] (asis) {};
        \node[
            rectangle, 
            draw=lightgreenn, 
            inner sep=15pt, 
            fit=(inflow2) (traffic2) (emissions2) (fake1) (fake2)
        ] (whatif) {};
    
    \draw[-Stealth, thick] (asis) -- (m1);
    \draw[-Stealth, thick] (m1) -- (inflow2);
    \draw[-Stealth, thick] (inflow2) -- (m2);
    \draw[-Stealth, thick] (m2) -- (traffic2);
    \draw[-Stealth, thick] (traffic2) -- (m3);
    \draw[-Stealth, thick] (m3) -- (emissions2);
    \draw[-Stealth, thick] (control.north) -- (m1.south);
    \draw[-Stealth, dashed] (control.north) -- (m2.south);
    \draw[-Stealth, thick] (control.north) -- (m3.south);
    \draw[-Stealth, thick] (behavioral.north) -- (m1.south);
    \draw[-Stealth, thick] (behavioral.north) -- (m2.south);
    \draw[-Stealth, dashed] (behavioral.north) -- (m3.south);
    \draw[-Stealth] (m1.east) -- (oii1);
    \draw[-Stealth] (m2.east) -- (oii2);
    \draw[-Stealth] (m3.east) -- (oii3);

\end{tikzpicture}
}
    \caption{Conceptual diagram of the case study, showing the main quantities of interest (i.e., inflow, traffic, and emissions) in the as-id and what-if scenarios. Blue squares represent modeling steps that map inputs to outputs, allowing for the assessment of the modified quantities.}
    \label{fig:scenario}
\end{figure}